\documentclass[a4paper,11pt]{article}
\usepackage{pst-3d}
\usepackage{pst-3dplot}
\usepackage[utf8]{inputenc}
\usepackage[english]{babel}
\usepackage{amsmath,amssymb,amsthm,mathrsfs,amsfonts,dsfont}
\usepackage{graphicx}
\usepackage[footnotesize, skip=0pt]{caption}
\usepackage{epigraph}
\usepackage{indentfirst}
\usepackage{booktabs}
\usepackage{fancyhdr}
\usepackage{vmargin}
\usepackage{lscape}
\usepackage{subfig}
\usepackage{textcomp}
\usepackage{pstricks}
\usepackage[metapost]{mfpic}
\usepackage{fancybox}
\usepackage{slashed}
\usepackage[verbose]{wrapfig}
\usepackage{pifont}
\usepackage{keystroke}
\usepackage[title]{appendix}
\usepackage{enumitem}
\usepackage{array}
\usepackage{colortbl}
\usepackage{alltt}
\usepackage{boxedminipage}
\usepackage{tikz}
\usepackage{calc}
\usepackage{multirow}

\usepackage{cmll}
\usepackage{multicol}
\definecolor{color1}{RGB}{204,0,51}
\definecolor{color2}{RGB}{159,182,205}
\usepackage[open,openlevel=3]{bookmark}

\usepackage{verbatim}
\usepackage{pst-grad} 
\usepackage{pst-plot}
\usepackage{transparent}
\usepackage{color,soul}
\usepackage{cancel}
\usepackage{placeins}
\usepackage{mathtools}
\usepackage{rotating}
\usepackage[refpage]{nomencl}
\usepackage{hhline}
\usepackage{marginnote}
\usepackage{upgreek}
\usepackage{listings}
\usepackage{hyperref}
\usepackage{cleveref}
\usepackage{afterpage}
\usepackage{cite}
\usepackage{makecell}
\usepackage{accents}
\usepackage{breqn}
\usepackage{environ}
\usepackage{empheq}
\usepackage{bbding}
\usepackage{titlesec}
\usepackage[final]{showlabels}
\usepackage{orcidlink}
\usepackage{nicematrix}

\newcommand{\CA}{\alpha}
\newcommand{\CB}{\beta}
\newcommand{\CC}{\gamma}

\newcommand{\PA}{\texttt{A}}
\newcommand{\PB}{\texttt{B}}
\newcommand{\PC}{\texttt{C}}
\newcommand{\PD}{\texttt{D}}

\newcommand{\BA}{\textcolor{green!90!black}{\texttt{a}}}
\newcommand{\BB}{\textcolor{green!90!black}{\texttt{b}}}

\newcommand{\GammaNR}{\tilde{\Gamma}}
\newcommand{\RiemNR}{\tilde{\rm R}}
\newcommand{\RicNR}{\tilde{\rm R}}

\newcommand{\nablaNR}{\tilde{\nabla}}

\newcommand{\Ct}{\tilde{\nabla}(t)}
\newcommand{\Cz}{\tilde{\nabla}(z)}
\newcommand{\Cm}{\tilde{\nabla}(f)}

\newcommand{\cc}[1]{\textcolor{ccolor}{c^{#1}}}
\definecolor{ccolor}{RGB}{255, 95, 31}
\definecolor{cdolor}{RGB}{255, 0, 0}

\newcommand{\R}{{\rm R}}
\newcommand{\Ric}{{\rm R}}
\newcommand{\Riem}{{\rm R}}

\newcommand{\oJ}{\omega}
\newcommand{\oG}{\omega}

\newcommand{\lorentzp}{\lambda}
\newcommand{\boostp}{\lambda}

\newcommand{\RH}{{\rm R(H)}}
\newcommand{\RM}{{\rm R(M)}}
\newcommand{\RP}{{\rm R(P)}}
\newcommand{\RG}{{\rm R(G)}}
\newcommand{\RJ}{{\rm R(J)}}
\newcommand{\RicJ}{{\rm Ric(J)}}
\newcommand{\CDRH}{{\tilde{\nabla} \rm R(H)}}

\allowdisplaybreaks

\hypersetup{
    bookmarks=false,         
    unicode=false,          
    pdftoolbar=true,        
    pdfmenubar=true,        
    pdffitwindow=false,     
    pdfstartview={FitH},    
    pdftitle={Higher-Order Newton-Cartan Gravity},    
    pdfauthor={Biel Cardona, Luca Romano},     
    pdfsubject={},   
    pdfnewwindow=true,      
    colorlinks=false,       
    linkcolor=red,          
    citecolor=red,        
    filecolor=red,      
    urlcolor=red           
    linkbordercolor={red},
    citebordercolor={red},
    urlbordercolor={red},
    bookmarksopen={true},
    linktocpage=true
}

\usepackage{listings} 

\makeatletter

\newcommand{\Rmnum}[1]{\expandafter\@slowromancap\romannumeral #1@}
\makeatother

\usepackage{alphalph}

\makeatletter
\newalphalph{\aalphalph}[mult]{\alphalph@alph}{26}
\newcommand{\alphalphval}[1]{%
  \@ifundefined{c@#1}{
    \aalphalph{#1}
  }{%
    \aalphalph{\value{#1}}
  }
}
\makeatother

\usepackage{color}

\def\chapterautorefname~#1\null{Chap.~(#1)\null}
\def\sectionautorefname~#1\null{Sec.~(#1)\null}
\def\subsectionautorefname~#1\null{sub--Sec.~(#1)\null}
\def\figureautorefname~#1\null{Fig.~(#1)\null}
\def\tableautorefname~#1\null{Tab.~(#1)\null}
\def\equationautorefname~#1\null{~(#1)\null}

\newcommand{\Autoref}[1]{%
  \begingroup%
  \def\chapterautorefname~##1\null{Chapter~(##1)\null}%
  \def\sectionautorefname~##1\null{Section~##1\null}%
  \def\subsectionautorefname~##1\null{Sub--Section~(##1)\null}%
  \def\figureautorefname~##1\null{Figure~(##1)\null}%
  \def\tableautorefname~##1\null{Table~(##1)\null}%
  \def\equationautorefname~##1\null{Equation~(##1)\null}%
  \autoref{#1}%
  \endgroup%
}
\def\equationautorefname~#1\null{eq.~(#1)\null}

\DeclareMathAlphabet\mathbfcal{OMS}{cmsy}{b}{n}

\usepackage{datetime}
\title{\huge\bf Higher-Order Newton-Cartan Gravity}

\date{}

\newcommand{\myalphfoot}
{
\renewcommand{\thefootnote}{\alph{footnote}}
}

\author{
\myalphfoot
{\bf\large Biel~Cardona\,\,\orcidlink{0000-0002-2943-8685}$^{~1}$}\footnotemark[1]\,\,
{\bf and}
{\bf\large Luca~Romano\,\,\orcidlink{0000-0001-9033-1345}$^{~1}$}\footnotemark[2]\,\,
\setcounter{footnote}{0}
\renewcommand{\thefootnote}{\arabic{footnote}}
}

\renewcommand*{\thefootnote}{\alph{footnote}}
\footnotetext[1]{Email: {\tt  cardonarotger@gmail.com}}
\footnotetext[2]{Email: {\tt lucaromano2607@gmail.com}}
\renewcommand{\thefootnote}{\arabic{footnote}}

\begin{document}

\begin{flushright}
\small
\today
\normalsize
\end{flushright}
{\let\newpage\relax\maketitle}
\maketitle
\def\equationautorefname~#1\null{(#1)\null}
\def\tableautorefname~#1\null{tab.~(#1)\null}
\def\sectionautorefname~#1\null{section #1\null}
\def\subsectionautorefname~#1\null{section #1\null}
\def\subsubsectionautorefname~#1\null{section #1\null}
\def\appendixautorefname~#1\null{appendix #1\null}
\def\figureautorefname~#1\null{figure~#1\null}

\vspace{0.8cm}

\begin{center}

\vspace{0.5cm}
${}^1${\it Departamento de Electromagnetismo y Electrónica, Universidad de Murcia,\\ Campus de Espinardo, 30100 Murcia, Spain }\\

\vspace{1.8cm}


{\bf Abstract}
\end{center}
\begin{quotation}
{\small
\noindent We study the non-relativistic Newton-Cartan limit of higher-order gravity theories in arbitrary dimensions. We first study it at the level of the action by introducing an additional 1-form gauge field and coupling it appropriately to the gravity sector. We extend this procedure to any theory whose Lagrangian is a function of the Ricci scalar, quadratic Ricci tensor and quadratic Riemann tensor. We also study the limit of the equations of motion for two models, Einstein-Gauss-Bonnet gravity and quadratic Ricci scalar theory. We prove that, in the first case, it is possible to obtain the Poisson equation by introducing a scalar field and imposing an appropriate constraint.
In the latter case, we show that it is possible to get the Poisson equation from the limit of the equations of motion as long as the on-shell constraint used in the two-derivative theory is supplemented with a further condition. We give the expressions of the two higher-order corrected Poisson equations in terms of curvatures of Newton-Cartan geometry. In both cases, we derive the full set of non-relativistic equations and study their boost transformations. The two sets of equations of motion define zero-torsion Gauss-Bonnet Newton-Cartan gravity and zero-torsion quadratic Ricci scalar Newton-Cartan gravity. }
\end{quotation}

\newpage

\tableofcontents

\section*{Introduction}\addcontentsline{toc}{section}{\protect\numberline{}Introduction}
\noindent The generalised correspondence principle is a cornerstone principle in theoretical physics. It asserts that a new theory, aiming to explain phenomena not addressed by previous ones, should, in certain limits or regimes, reproduce the established results of well-tested theories. This principle underpins the robustness of our knowledge and it can be naturally used as a theoretical test for new theories. Non-Lorentzian theories obtained via a limit of Lorentzian ones provide compelling examples in which inspecting a specific regime gives both a consistency check of the parent theory and possible new insights.

Among all non-Lorentzian theories a prominent role is held by Newton-Cartan gravity \cite{ASENS_1924_3_41__1_0, ASENS_1923_3_40__325_0, Friedrichs1928EineIF}, the covariant formulation of Newtonian gravity. 
Newton-Cartan geometry stands for Newton-Cartan gravity as Riemannian geometry stands for General Relativity. This non-Lorentzian geometry, in contrast with Riemannian geometry, is characterised by the presence of two degenerate metrics. The fields of Newton-Cartan gravity are a clock form $\tau_{\mu}$, a spatial vielbein $e_{\mu}{}^{a}$ and a one-form gauge field $a_{\mu}$, associated with the conservation of mass \cite{Bergshoeff:2022fzb, Bergshoeff:2022eog, Hartong:2022lsy}.  Analogously to General Relativity, which can be obtained by gauging the Poincaré algebra, Newton-Cartan gravity can be built by gauging the Bargmann algebra, the central extension of the Galilei algebra \cite{Duval:1984cj, Andringa:2010it}.

Another way to obtain Newton-Cartan gravity is via a non-relativistic limit of General Relativity \cite{Bergshoeff:2015uaa}. The limit is a procedure based on two main steps. The first is to redefine the relativistic fields in terms of the would-be non-relativistic ones, involving a parameter. The second step consists of sending the parameter to a critical value. In the present work this parameter can be identified with the speed of light, which we denote by $\cc{}$ and it is sent to infinity to inspect the non-relativistic regime. The construction of Newton-Cartan gravity from a relativistic theory is based on a decomposition of the spacetime tangent space into a time-like direction and the remaining space-like directions. This decomposition, in $D$ dimensions, induces the breaking of the Lorentz symmetry as
\begin{align*}
SO(1,D-1)\rightarrow SO(D-1)\ltimes \mathds{R}^{D-1}\,,
\end{align*}
where the last factor corresponds to the Galilean boosts. This breaking pattern has been generalised to
\begin{align*}
SO(1,D-1)\rightarrow SO(1,p)\times SO(D-p-1)\ltimes \mathds{R}^{D-p-1}\,,
\end{align*}
to accommodate extended objects, such as strings or $p$-branes. The limits defined by this decomposition are called $p$-brane limits \cite{Andringa:2010it, Andringa:2012uz}. These generalisations have proven useful in studying the limit of string theories and supergravity \cite{Gomis:2000bd, Bergshoeff:2023fcf, Bergshoeff:2021tfn, Blair:2021waq, Bergshoeff:2023ogz, Bergshoeff:2024nin, Bergshoeff:2025grj, Bergshoeff:2021bmc, Andringa:2013mma, 3823ef0bb56b497c9abff0dec317478b, f5820c1fd6464ce49de6ce759b35bb1a, Bergshoeff:2016lwr, Bergshoeff:2022iyb}.

The recent surge of interest in non-Lorentzian theories and geometries has been driven by their relevance in various contexts. In the non-relativistic regime, Newton-Cartan geometry has been used to write an effective action for the quantum Hall effect \cite{Son:2013rqa, Geracie:2014nka}, to investigate Hall viscosity \cite{Copetti:2019rfp} in condensed matter and in Lifshitz holography \cite{Christensen:2013lma, Christensen:2013rfa}. On the dual side, Carroll symmetry, the ultra-relativistic analogue of Galilean symmetry in the ultra-relativistic regime \cite{Hartong:2015xda} could explain the vanishing of the Love numbers of four-dimensional Schwarzschild black holes \cite{Penna:2018gfx}. Carrollian physics plays a crucial role in describing the near-horizon geometry of black holes, in the BMS group, as well as in holography \cite{Bergshoeff:2017btm, Donnay:2019jiz, Duval:2014uva, Donnay:2022aba, Blair:2025nno}. It could also be relevant in the context of dark energy and inflation \cite{deBoer:2021jej}.

One further motivation to investigate the non-Lorentzian regime resides in the possibility of isolating corners of relativistic theories that can provide us with useful information to build a non-Lorentzian quantum gravity theory. In an alternative approach to the search for a quantum gravity theory, this would be an intermediate step towards it, the final step being the inclusion of relativistic corrections. In this regard, it is natural to study the different aspects and features of string theory that could serve as a probe to non-perturbative regimes, like BPS brane solutions and higher-order corrections to General Relativity, and test them by performing a non-Lorentzian limit \cite{Bergshoeff:2022pzk, Oling:2022fft, Blair:2021waq, Blair:2024aqz, Blair:2023noj, Cardona:2016ytk}. Higher-order corrections arise naturally in String Theory, which generally predicts an infinite number of such subleading corrections to the Einstein-Hilbert  action\cite{Gross:1986iv, Grisaru:1986vi, Gross:1986mw}.

In light of these considerations we want to investigate the non-relativistic regime of higher-order gravity theories. Pioneering results have been obtained by \cite{Tadros:2023teq, Tadros:2024fgi, Tadros:2024qlo, Lescano:2025yio}. Aiming to add a piece of knowledge to this line of research, in the present paper, we consider limits with a symmetry given by the Bargmann algebra, in an attempt to obtain higher-order Newton-Cartan gravity.  We furthermore investigate the limit both at the level of the action and equations of motion with a focus on the possibility of obtaining the higher-order corrected Poisson equation.

The paper is organized as follows. In \autoref{sec:RelaHDT} we briefly introduce the relativistic quadratic-order theories we consider in this work. In \autoref{sec:NRActionLimit} we first review the electric and magnetic limits of the Einstein-Hilbert action and then we study the non-relativistic limit of higher-order gravity theories at the level of the action. We are interested in Newton-Cartan type of limits, so we use a one-form gauge field, suitably coupled to the gravity sector, to cancel the ``divergent'' orders of the $\cc{}$-expansion of the action. We provide a recipe to build a well-defined non-relativistic limit for any theory whose action is a function of the Ricci scalar, the squared Ricci tensor and the squared Riemann tensor. \Autoref{sec:EomLimit} is devoted to the analysis of the non-relativistic limit at the level of the equations of motion, with a particular emphasis on obtaining the Poisson equation from the limit. After reviewing how it is possible to get the Poisson equation in the two-derivative case, we generalise the procedure to higher-order theories. In \autoref{sec:GBcase} we consider the Einstein-Gauss-Bonnet theory and we show that, by slightly deforming the initial model with the help of a scalar field in \autoref{sec:SclarFieldTrick} it is possible to obtain the Poisson equation.  We derive the full set of equations in \autoref{sec:NRGBEOM}
and prove that they form a non-trivial closed set, organized in a decomposable irreducible representation of the Bargmann algebra. This set of equations describes zero-torsion Gauss-Bonnet Newton-Cartan gravity.  In \autoref{sec:HigherOrderEOM} we repeat the same analysis for the quadratic Ricci scalar theory.  We show that it is possible to derive the Poisson equation by supplementing the system with an on-shell constraint. This adds up to that considered in the two-derivative case, whose limit leads to zero intrinsic torsion. We close the section by deriving the full set of non-relativistic equations of motion and their transformations under boost \autoref{sec:NRRREOM}.  The system defines zero-torsion quadratic Ricci Newton-Cartan gravity.  We conclude the paper with some final remarks and perspectives.  We have three appendices, \autoref{sec:Notation}, \autoref{sec:UsRel}, and  \autoref{sec:Geometry}, that summarize the notation, some useful relations and conventions and the Newton-Cartan geometric quantities, respectively.

Most of the calculations have been carried out with the help of Cadabra \cite{Peeters:2007wn, Peeters2018, Peeters:2006kp} and xAct for Mathematica \cite{xAct, Mathematica}.  

\section{Relativistic Quadratic Gravity Theories}\label{sec:RelaHDT}
\noindent In this section we introduce the relativistic theories whose non-relativistic limit will be studied later. We are interested in actions that are at most quadratic in the curvatures. These include up to the quartic power of the derivative of the metric. In particular, we consider theories described by the following action (we refer the interested reader to \cite{Alvarez-Gaume:2015rwa, Donoghue:2021cza, Salvio:2018crh} for a more detailed analysis of these theories): 
\begin{align}
\mathcal{S}=&\,  \int d^{D}x\, \sqrt{-g}\,\Big( \R + \CA \, \R^{2} + \CB\, \Ric_{\mu\nu}\Ric^{\mu\nu} + \CC\,  \Riem_{\mu\nu\rho\sigma}\Riem^{\mu\nu\rho\sigma}\Big)\,,\label{eq:quadraticgravity}
\end{align}
where $g$ denotes the determinant of the spacetime metric $g_{\mu\nu}$; $\Riem_{\mu\nu\rho\sigma}$, $\Ric_{\mu\nu}$ and $\R$ are the Riemann tensor, Ricci tensor and Ricci scalar respectively and $\CA,\CB$ and $\CC$ are three coupling constants. Definitions of the fundamental quantities are listed in \autoref{sec:UsRel}. We refer to \autoref{sec:Notation} for details about the notation. We work in $D$ dimensions and we adopt mostly-plus signature for the tangent space metric. We do not assume the higher-derivative terms to be perturbative corrections, i.e. we do not assume $\CA,\CB$ and $\CC$ to be small. We consider these as non-perturbative corrections.

Among the purely quadratic theories parameterised by different values of $\CA,\CB$ and $\CC$ there is one theory that enjoys special properties, this is Gauss-Bonnet gravity. Gauss-Bonnet gravity is the theory described by setting $\CB=-4\CA$ and $\CC=\CA$ in \autoref{eq:quadraticgravity}. It is a topological invariant in four dimensions and despite having an action that is fourth-order in the derivative of the metric the equations of motion are second-order. We investigate the non-relativistic limit of the equations of motion of Einstein-Gauss-Bonnet gravity, i.e. the Einstein-Hilbert action supplemented with the Gauss-Bonnet term, in \autoref{sec:GBcase}, where we also find the corresponding Poisson equation. We repeat the same analysis for the quadratic Ricci scalar gravity in \autoref{sec:HigherOrderEOM}, i.e. the theory defined by $\CB=\CC=0$ in \autoref{eq:quadraticgravity}.

In the next section we study the non-relativistic limit of the action \autoref{eq:quadraticgravity}. The procedure extends to any theory whose Lagrangian is written as a function of the Ricci scalar,  $\Ric_{\mu\nu}\Ric^{\mu\nu}$ and   $\Riem_{\mu\nu\rho\sigma}\Riem^{\mu\nu\rho\sigma}$.

\section{Non-Relativistic Limit of the Action}\label{sec:NRActionLimit}
\noindent In this section we study the magnetic limit of the higher-order theories introduced in \autoref{sec:RelaHDT} at the level of the action. We start by reviewing the electric and magnetic limits of the Einstein-Hilbert-Maxwell action and then we extend the procedure to relativistic gravity theory with higher-order corrections. 

\subsection{Electric \& Magnetic Limits of the Einstein-Hilbert-Maxwell Action}
\noindent To study the theories described in \autoref{sec:RelaHDT} it is useful first to briefly review the two-derivative case of the Einstein-Hilbert action. We denote with $E_{\mu}{}^{\hat{A}}$ the $D$-dimensional vielbein. The $D$-dimensional flat index splits as $\hat{A}=\{0,a\}$ where $a=1,..., D-1$. There are two main possibilities to take the limit of the Einstein-Hilbert action. The first is to define an ansatz for the metric, expand the action plugging in it this ansatz, and send $\cc{}$ to infinity. This procedure will select the highest power of the expansion of the action. This approach is known as the {\it electric} limit. We briefly review how it works. We consider the following ansatz for the $D$-dimensional vielbein
\begin{subequations}
\begin{align}
E_{\mu}{}^{0}=&\,  \cc{}\tau_{\mu}\,,& E^{\mu}{}_{0}=&\,  \frac{1}{\cc{}}\tau^{\mu}\,,\\
E_{\mu}{}^{a}=&\,  e_{\mu}{}^{a}\,,& E^{\mu}{}_{a}=&\,  e_{\mu}{}_{a}\,.
\end{align}\label{eq:ansatze}
\end{subequations}
The would-be non-relativistic fields $\tau_{\mu}$, $e_{\mu}{}^{a}$ and their inverses satisfy the following relations:
\begin{subequations}
\begin{align}
e_{\mu}{}^{a}e^{\mu}{}_{b}=&\,  \delta_{b}^{a}\,,&\tau_{\mu}\tau^{\mu}=&\,  1\,,\\
e_{\mu}{}^{a}\tau^{\mu}=\tau_{\mu}e^{\mu}{}_{b}=&\,  0\,,&\tau_{\mu}\tau^{\nu}+e_{\mu}{}^{a}e^{\nu}{}_{a}=&\,  \delta_{\mu}^{\nu}\,.
\end{align}
\end{subequations}
The ansatz \autoref{eq:ansatze} implies the following expansion of the metric and its inverse
\begin{subequations}
\begin{align}
g_{\mu\nu}=&\,  E_{\mu}{}^{\hat{A}}E_{\nu}{}^{\hat{B}}\eta_{\hat{A}\hat{B}}=-\cc{2}\tau_{\mu}\tau_{\nu}+h_{\mu\nu}\,,\\
g^{\mu\nu}=&\,  E^{\mu}{}_{\hat{A}}E^{\nu}{}_{\hat{B}}\eta^{\hat{A}\hat{B}}=-\frac{1}{\cc{2}}\tau^{\mu}\tau^{\nu}+h^{\mu\nu}\,.
\end{align}
\end{subequations}
Plugging the ansatz into the Ricci scalar we get the following expansion:
\begin{align}
\mathcal{S}=&\,    \frac{\cc{2}}{4}\int d^{D}x\, e\, t_{ab}t^{ab}+\mathcal{O}(\cc{0})\,,\label{eq:EHexpanded}
\end{align}
where $t_{ab}$ is the transverse component of the intrinsic torsion, defined by
\begin{align}
t_{ab}=&\,  e^{\mu}{}_{a}e^{\nu}{}_{b}\, 2\partial_{[\mu}\tau_{\nu]}\,,
\end{align}
(see \autoref{sec:Notation} for further details) and 
\begin{align}
e=\det (\tau_{\mu}\,, e_{\mu}{}^{a})=&\,  \epsilon^{\mu_{1}...\mu_{D}}\tau_{\mu_{1}}e_{\mu_{2}}{}^{1}...e_{\mu_{D}}{}^{D-1}\,.
\end{align}
We have absorbed the power of $\cc{}$ coming from the determinant in an overall rescaling and we will do so also in the next. Performing the limit $\cc{}\rightarrow \infty$ in \autoref{eq:EHexpanded} amounts to rescaling the expression by an overall power $\cc{-2}$ and selecting then the leading order. The result is simply \autoref{eq:EHexpanded} without the $\cc{2}$ factor. 

An alternative approach is to regard the leading term of \autoref{eq:EHexpanded} as a divergent term and try to find a way to cancel it in such a way as to get access to the subleading order of the expansion. We refer to this as the {\it magnetic} limit. To achieve the cancellation of the leading order term in \autoref{eq:EHexpanded} we modify our starting point by adding a Maxwell term in the action 
\begin{align}
\mathcal{S}=&\,  \int d^{D}x\sqrt{-g}\,\bigg( \R-\frac{1}{4}F_{\mu\nu}F^{\mu\nu}\bigg)\,,
\end{align}
where $F_{\mu\nu}=2\partial_{[\mu}A_{\nu]}$ and $A_{\mu}$ is a 1-form gauge field. By choosing the ansatz for the gauge field one tries to cancel the leading order term coming from the Ricci scalar.  In particular, we consider the following ansatz:
\begin{align}
A_{\mu}=&\,  \cc{}\tau_{\mu}+\frac{1}{\cc{}}a_{\mu}\,,
\end{align}
with $a_{\mu}$ playing the role of the non-relativistic gauge field after the limit. With this, the field strength expands as
\begin{align}
F_{\mu\nu}=&\,  \cc{}t_{\mu\nu}+\frac{1}{\cc{}}f_{\mu\nu}\,,
\end{align}
where $f_{\mu\nu}=2\partial_{[\mu}a_{\nu]}$. Then the expansion of the action reads
\begin{align}
\mathcal{S}=&\,    \int d^{D}x\, e\, \bigg( e^{\mu a} e^{\nu}_{a} \RicNR_{\mu \nu} + 2\tau^{\mu} e^{\nu a} e^{\rho}_{a} \nablaNR_{\nu}t_{\mu \rho}  -  \frac{3}{2} t_{0 a} t_{0}{}^{a} - t^{a b} f_{a b}\bigg)+\mathcal{O}(\cc{-2})\,,\label{eq:EHexpanded2}
\end{align}
with $\RicNR_{\mu\nu}$ being a tensor defined using the following connection
\begin{align}
\GammaNR_{\mu\nu}^{\rho}=&\,  \frac{1}{2}h^{\rho\xi}(\partial_{\mu}h_{\nu\xi}+\partial_{\nu}h_{\mu\xi}-\partial_{\xi}h_{\mu\nu})+\tau^{\rho}\partial_{\mu}\tau_{\nu}+h^{\rho\xi}\tau_{(\mu}f_{\nu)\xi}\,,\label{eq:NRConnection}
\end{align}
that mimics the definition of the relativistic Ricci tensor, see \autoref{sec:UsRel}. The connection \autoref{eq:NRConnection} is such that
\begin{align}
\nablaNR_{\mu}\tau_{\nu}=&\,  0\,,& \nablaNR_{\mu}h^{\nu\rho}=&\,  0\,,
\end{align}
where $\nablaNR_{\mu}$ denotes the covariant derivative containing only the non-relativistic connection defined above. This connection is boost-covariant and gauge invariant. See \autoref{sec:Geometry} for further details. 

In the limit $\cc{}\rightarrow \infty$ the term of order $\cc{0}$ in \autoref{eq:EHexpanded2} is the only surviving term. This is the result of the magnetic limit. We can rewrite \autoref{eq:EHexpanded2} conveniently in term of the Newton-Cartan geometric quantities defined in \autoref{sec:Geometry} as
\begin{align}
\mathcal{S}=&\,    \int d^{D}x\, e\, \bigg[-\RicJ+2\tau^{\mu}h^{\nu\rho}\nablaNR_{\nu}\RH_{\mu\rho}-\frac{3}{2}\RH_{0a}\RH_{0}{}^{a} \bigg]\, \,.
\end{align}

Before moving our attention to the higher-order case it is useful to report the limit of the Lorentz and gauge transformation rules, 
\begin{subequations}
\begin{align}
\delta \tau_{\mu}=&\,  0\,,\\
\delta e_{\mu}{}^{a}=&\,  \lambda^{a}{}_{b}e_{\mu}{}^{b}+\lambda^{a}\tau_{\mu}\,,\\
\delta e^{\mu}{}_{a}=&\,  \lambda_{a}{}^{b}e^{\mu}{}_{b}\,,\\
\delta \tau^{\mu}=&\,  -\lambda^{b}e^{\mu}{}_{b}\,,\\
\delta a_{\mu}=&\,  \partial_{\mu}\sigma -\lambda_{a}e_{\mu}{}^{a}\,,
\end{align}\label{eq:BargmannAlgebra}
\end{subequations}
obtained by adopting the following ansatz for the relativistic parameters
\begin{align}
\Lambda_{ab} =&\,  \lambda_{ab}\,,& \Lambda_{0a} =&\,  -\frac{1}{\cc{}}\lambda_{a}\,,& \Sigma =&\,  \frac{1}{\cc{}}\sigma\,.
\end{align}
We do not report the diffeomorphisms, since they are unaffected by the limit.  The transformations \autoref{eq:BargmannAlgebra} are those obtained by a gauging of the Bargmann algebra, the central extension of the Galilei algebra. The introduction of the Maxwell term in the relativistic theory provides the extra U(1) generator useful to land on the Bargmann algebra.

\subsection{Magnetic Limit of Higher-Order Gravity Theories}
\noindent We would like to extend the procedure to take the magnetic limit to the case of higher-order gravity theories. Our starting point is the action introduced in \autoref{sec:RelaHDT}.  We will study the three different terms separately. In taking the limit the couplings will not be rescaled by any power of $\cc{}$. In analogy with the two-derivative case, we expect that, to take the magnetic limit, we should introduce not only the two-derivative Maxwell term but also four-derivative Maxwell terms and four-derivative couplings between the gauge field and the gravity sector. Since we already know that the two-derivative Einstein-Hilbert-Maxwell finite part 
occurs at order $\cc{0}$ we should require all the orders higher than that in the expansion of the higher-derivative terms to cancel. If this is not the case then the contribution to the limit of the two-derivative part would be washed out after the limit. In general, it emerges that achieving a full cancellation of all the orders greater than $\cc{0}$ is harder than the two-derivative case. In particular, the four-derivative terms produce non-trivial diverging terms both at order $\cc{4}$ and $\cc{2}$ as follows:
\begin{subequations}
\begin{align}
\R^2=&\,  \frac{1}{16}\cc{4}t_{ab}t^{ab}t_{cd}t^{cd}+\nonumber\\
&+\frac{1}{4}\cc{2}t_{ab}t^{ab}\Big(2\RicNR_{c}{}^{c}+4\Ct_{c0}{}^{c}-4t_{0c}t_{0}{}^{c}-t^{cd}f_{cd}\Big)+\mathcal{O}(\cc{0})\,,\\
\nonumber\\
\Ric_{\mu\nu}\Ric^{\mu\nu}=&\,  \frac{1}{16}\cc{4}\Big(t_{ab}t^{ab}t_{cd}t^{cd}+4t_{a}{}^{b}t_{b}{}^{c}t_{c}{}^{d}t_{d}{}^{a}\Big)+\,,\nonumber\\
&+\frac{1}{4}\cc{2}\Big[4t^{ab}t_{a}{}^{c}(\RicNR_{bc}+\Ct_{b0c}-3t_{0b}t_{0c})+\nonumber\\
&+t^{ab}t_{ab}(-2\Ct_{c0}{}^{c}+2t_{0c}t_{0}{}^{c}+t^{cd}f_{cd})+\nonumber\\
&+2(\Ct_{a}{}^{ab}-4t^{ab}t_{0a})\Ct_{cb}{}^{b}\Big]+\mathcal{O}(\cc{0})\,,\\\nonumber\\
\Riem_{\mu\nu\rho\sigma}\Riem^{\mu\nu\rho\sigma}=&\,  \frac{1}{8}\cc{4}\Big(3t_{ab}t^{ab}t_{cd}t^{cd}+5t_{a}{}^{b}t_{b}{}^{c}t_{c}{}^{d}t_{d}{}^{a}\Big)+\nonumber\\
&+\cc{2}\bigg[2t^{ab}t^{cd}\bigg(\RiemNR_{a(bc)d}+\frac{1}{4}t_{ab}f_{cd}\bigg)-\frac{4}{3}\Ct^{abc}\Ct_{(ab)c}+\nonumber\\
&-2t^{ab}t_{a}{}^{c}\bigg(\Ct_{b0c}+\frac{1}{3}t_{0b}t_{0c}-\frac{3}{4}t_{b}{}^{d}f_{cd}\bigg)+\nonumber\\
&+\frac{8}{3}t^{ab}t_{0}{}^{c}(2\Ct_{(ac)b}-t_{ab}t_{0c})\bigg]+\mathcal{O}(\cc{0})\,,
\end{align}
\end{subequations}
where $\Ct_{\mu\nu\rho}:=\nablaNR_{\mu}t_{\nu\rho}$. We start our analysis by considering the Ricci scalar squared term. For this term the knowledge acquired from the two-derivative case suggests us to modify it as
\begin{align}
\mathcal{L}_{\CA}=&\,  \bigg(\R-\frac{1}{4}F_{\mu\nu}F^{\mu\nu}\bigg)^{2}\,.
\end{align}
For the other terms the situation is more involved so we can define a more systematic approach. We consider a basis of possible four-derivative terms involving the Maxwell tensor and curvatures. The basis is:
\begin{align}
\mathcal{B}=\Big\{&(F_{\mu\nu}F^{\mu\nu})^2\,\,,  F^{\mu}{}_{\nu}F^{\nu}{}_{\rho}F^{\rho}{}_{\sigma}F^{\sigma}{}_{\mu}\,,F^{\mu\nu}F^{\rho\sigma}\Riem_{\mu\nu\rho\sigma}\,, F^{\mu\rho}F^{\nu}{}_{\rho}\Ric_{\mu\nu}\,,\nonumber\\
 &F_{\mu\nu}F^{\mu\nu}\R\,, \nabla_{\mu}F_{\nu\rho}\nabla^{\mu}F^{\nu\rho}\,, \nabla_{\mu}F^{\mu\rho}\nabla_{\nu}F^{\nu}{}_{\rho}\Big\}\,.
\end{align}
A key feature that distinguishes the two-derivative case from the four-derivative case and has an important impact in the present treatment is that, while in the two-derivative case the reciprocal coefficients between the Ricci scalar and the Maxwell term can be modified by a redefinition of the gauge field, keeping fixed the sign, in the presence of four-derivative terms the coefficients cannot be all arbitrarily changed by simple redefinitions. This is because the number of terms is usually greater than the number of fields that can be redefined. This makes the cancellation of the diverging terms harder to realize.

We proceed by adding the higher-derivative terms in $\mathcal{B}$ to the pure four-derivative gravitational part, parameterising each of them with a coefficient. We then fix the coefficients by requiring divergence cancellation. Although some of the action terms appearing in the basis above could be related via integration by parts, without loss of generality, we prefer to work with this set of terms because performing integration by parts when studying the cancellation of divergences can result in more involved calculations. Thus we have considered all the independent terms up to Bianchi identities and terms of the type $F\nabla\nabla F$, since it can be proved that these do not play any role in the cancellation of divergences.
By proceeding as described we find in all three cases one single solution. More precisely, for each of the three quadratic terms in \autoref{eq:quadraticgravity} there is one single combination of the terms in $\mathcal{B}$ that cancels the corresponding divergent contribution. The first case being the one we have already predicted. Then \autoref{eq:quadraticgravity} should modified as
\begin{align}
\mathcal{S}=&\,  \int d^{D} x \sqrt{-g} \Big(\mathcal{L}_{ EHM}+\CA\mathcal{L}_{\CA}+\CB\mathcal{L}_{\CB}+\CC\mathcal{L}_{\CC}\Big)\,,
\end{align}
with $\mathcal{L}_{EHM}$ the Einstein-Hilbert-Maxwell Lagrangian and
\begin{subequations} 
\begin{align}
\mathcal{L}_{\CA}=&\,  \bigg(\R-\frac{1}{4}F_{\mu\nu}F^{\mu\nu}\bigg)^{2}\,,\label{eq:ActionsRelA}\\
\nonumber\\
\mathcal{L}_{\CB}=&\,    \Ric_{\mu\nu}\Ric^{\mu\nu}+\frac{1}{16}F^{4}+\frac{1}{4}F^{\mu}{}_{\nu}F^{\nu}{}_{\rho}F^{\rho}{}_{\sigma}F^{\sigma}{}_{\mu}-\frac{1}{2}\nabla_{\mu}F^{\mu\nu}\nabla_{\rho}F_{\nu}{}^{\rho}-F^{\mu\rho}F^{\nu}{}_{\rho}\Ric_{\mu\nu}\,,\label{eq:ActionsRelB}\\
\nonumber\\
\mathcal{L}_{\CC}=&\,  \Riem_{\mu\nu\rho\sigma}\Riem^{\mu\nu\rho\sigma}
+\frac{3}{8}F^{4}+\frac{5}{8}F^{\mu}{}_{\nu}F^{\nu}{}_{\rho}F^{\rho}{}_{\sigma}F^{\sigma}{}_{\mu}+\nabla_{\mu}F_{\nu\rho}\nabla^{\mu}F^{\nu\rho}-\frac{3}{2}F^{\mu\nu}F^{\rho\sigma}\Riem_{\mu\nu\rho\sigma}\,.\label{eq:ActionsRelC}
\end{align}
\label{eq:ActionsRel}
\end{subequations}
The cancellation of the diverging terms ensures that the non-relativistic limit is well-defined at the level of the action.  Additionally, the fact that the leading order of the Lagrangian terms \autoref{eq:ActionsRel} is $\cc{0}$ implies that the non-relativistic limit of any theory whose Lagrangian is a function of $\R$, $\Ric_{\mu\nu}\Ric^{\mu\nu}$ and $\Riem_{\mu\nu\rho\sigma}\Riem^{\mu\nu\rho\sigma}$,
\begin{align}
\mathcal{L}=&\,  f(R,\,\Ric_{\mu\nu}\Ric^{\mu\nu},\,\Riem_{\mu\nu\rho\sigma}\Riem^{\mu\nu\rho\sigma} )
\end{align}
is well defined upon the following substitution:   
\begin{align}
\mathcal{L}=&\,  f(\mathcal{L}_{EHM},\mathcal{L}_{\CB},\mathcal{L}_{\CC})\,.
\end{align}
In the next subsection we report the results of the non-relativistic limit of $\mathcal{L}_{\CA}$, $\mathcal{L}_{\CB}$ and $\mathcal{L}_{\CC}$. 

\subsubsection{Non-Relativistic Actions}
\noindent Performing the  limit $\cc{}\rightarrow \infty$ in \autoref{eq:ActionsRel} we find the following results:
\begin{subequations}
\begin{align}
\accentset{(0)}{\mathcal{L}}_{\alpha}=&\,  \bigg[\,\,\tilde{\rm R}_{a}{}^{a}+2\,\,\tilde{\nabla}(t)_{a0}{}^{a}-\frac{3}{2}t_{0a}t_{0}{}^{a}-f^{ab}t_{ab}\bigg]^{2}\,,\\
\nonumber\\
\accentset{(0)}{\mathcal{L}}_{\beta}=&\,  
\tilde{\rm R}_{ab}\Big[\tilde{\rm R}^{ab}-2\tilde{\nabla}(t)^{(ab)}{}_{0}-t_{0}{}^{a}t_{0}{}^{b}-2f^{c(a}t_{c}{}^{b)}+f^{c[a}t_{c}{}^{b]}+t^{c[a}z_{0c}{}^{b]}+t^{c[a}z_{0}{}^{b]}{}_{c}+t^{ab}z_{0c}{}^{c}\Big]+\nonumber\\
&+2\tilde{\rm R}_{(0b)}\Big(t^{ab}t_{0a}-\tilde{\nabla}(t)_{a}{}^{ab}\Big)+\frac{1}{2}t^{ab}t_{ab}\tilde{\rm R}_{00}+\tilde{\nabla}(t)_{a}{}^{ba}\Big[\tilde{\nabla}(t)_{00b}+f_{0}{}^{c}t_{cb}+f_{ab}t_{0}{}^{a}\Big]+\nonumber\\
&+\tilde{\nabla}(t)_{b0}{}^{b}\bigg[\frac{1}{2}\tilde{\nabla}(t)_{a0}{}^{a}-t_{0}{}^{a}t_{0a}\bigg]+t^{ab}t_{0a}\tilde{\nabla}(t)_{00b}+\tilde{\nabla}(t)_{ab0}\Big[\tilde{\nabla}(t)^{(ab)0}+t_{0}{}^{a}t_{0}{}^{b}+2f^{c(a}t_{c}{}^{b)}\Big]+\nonumber\\
&-f_{0}{}^{a}t_{a}{}^{b}t_{b}{}^{c}t_{0c}-\frac{1}{2}f_{0}{}^{a}t^{bc}t_{bc}t_{0a}+\frac{3}{4}t_{0}{}^{a}t_{0a}t_{0}{}^{b}t_{0b}+\frac{5}{8}f^{ab}f_{a}{}^{c}t_{b}{}^{d}t_{cd}+\frac{3}{8}f^{ab}f^{cd}t_{ac}t_{bd}+\nonumber\\
&+2f^{ab}t_{a}{}^{c}t_{0b}t_{0c}+\frac{1}{2}f^{ab}t_{a}{}^{c}t_{c}{}^{d}z_{0(bd)}+\frac{1}{4}t^{ab}t_{ab}z_{0c}{}^{c}z_{0d}{}^{d}-t^{ab}t_{a}{}^{c}z_{0bc}z_{0d}{}^{d}+\frac{1}{8}t^{ab}t_{a}{}^{c}z_{0b}{}^{d}z_{0cd}+\nonumber\\
&+\frac{1}{4}t^{ab}t_{a}{}^{c}z_{0b}{}^{d}z_{0dc}+\frac{1}{8}t^{ab}t_{a}{}^{c}z_{0}{}^{d}{}_{b}z_{0cd}+\frac{1}{2}t^{ab}t^{cd}z_{0ac}z_{0(bd)}\,,\\
\nonumber\\
\accentset{(0)}{\mathcal{L}}_{\gamma}=&\,  \frac{2}{3}\tilde{R}_{abcd}\bigg[2\tilde{R}^{a(bc)d}-f^{a(b}t^{c)d}+3f^{c(b}t^{d)a}+3t^{ac}z_{0}{}^{(bd)}+\frac{1}{2}t^{cb}z_{0}{}^{(ad)}+\frac{1}{2}t^{cd}z_{0}{}^{(ab)}\bigg]+\nonumber\\
&+\frac{2}{3}\tilde{R}_{ab0c}\Big[2\tilde{\nabla}(t)^{[ba]c}+2\tilde{\nabla}(t)^{cab}-t^{ab}t_{0}{}^{c}+t^{c[a}t_{0}{}^{b]}\Big]+\nonumber\\
&+\frac{2}{3}\tilde{R}_{a0bc}\Big[-2\tilde{\nabla}(t)^{(ab)c}+t^{b(c}t_{0}{}^{a)}\Big]-\tilde{R}_{b00c}t^{ab}t_{a}{}^{c}+\nonumber\\
&+\frac{2}{3}\tilde{\nabla}(t)^{abc}\Big[\tilde{\nabla}(m)_{(ab)c}-\tilde{\nabla}(z)_{0b(ac)}+\tilde{\nabla}(z)_{(ab)0c}-\tilde{\nabla}(z)_{b0(ac)}\Big]+\nonumber\\
&-\frac{2}{3}\tilde{\nabla}(t)_{0}{}^{ab}\Big[\tilde{\nabla}(t)_{0ab}+2\tilde{\nabla}(t)_{a0b}\Big]
+\frac{2}{3}\tilde{\nabla}(t)^{a}{}_{0}{}^{b}\Big[\tilde{\nabla}(t)_{a0b}+2\tilde{\nabla}(t)_{b0a}\Big]+\nonumber\\
&+\tilde{\nabla}(t)_{abc}\bigg[-4f_{0}{}^{(a}t^{b)c}+\frac{2}{3}t_{0}{}^{[a}z_{0}{}^{b]c}-\frac{2}{3}t_{0}{}^{b}z_{0}{}^{ca}+f^{db}z_{d}{}^{(ac)}-\frac{8}{3}f^{b(a}t_{0}{}^{c)}+z^{d(ab)}z_{0d}{}^{c}+z_{0}{}^{(c}{}_{d}z^{a)bd}\bigg]+\nonumber\\
&-2t_{0}{}^{a}t_{0}{}^{b}\tilde{\nabla}(t)_{a0b}+2t^{ab}t_{0a}\tilde{\nabla}(t)_{00b}-\frac{1}{3}t^{ab}t_{0}{}^{c}\tilde{\nabla}(m)_{(ac)b}+t^{ab}t_{a}{}^{c}\tilde{\nabla}(m)_{b0c}-t^{ab}t_{a}{}^{c}\tilde{\nabla}(z)_{00bc}+\nonumber\\
&+\frac{1}{3}t^{ab}t_{0}{}^{c}\Big[\tilde{\nabla}(z)_{a0cb}+\tilde{\nabla}(z)_{[c0]ab}+\tilde{\nabla}(z)_{(a0)bc}\Big]+\nonumber\\
&+\frac{3}{4}t_{0}{}^{a}t_{0a}t_{0}{}^{b}t_{0b}-\frac{4}{3}f^{ab}t_{0}{}^{c}t_{0(c}t_{b)a}+\frac{1}{12}f^{ab}t^{cd}(10f_{a(b}t_{c)d}+3t_{ab}f_{cd})+\nonumber\\
&+f^{ab}t_{a}{}^{c}\bigg[\frac{1}{2}z_{b(dc)}t_{0}{}^{d}+\frac{1}{2}t_{c}{}^{d}z_{0bd}+\frac{4}{3}t_{c}{}^{d}z_{0(bd)}\bigg]-f_{0}{}^{a}t^{bc}t_{b}{}^{d}z_{acd}+\frac{1}{3}t^{ab}t_{0}{}^{c}(t_{0a}z_{0bc}+t_{0(a|}z_{0|c)b})+\nonumber\\
&+\frac{2}{3}t^{ab}z_{0}{}^{cd}t_{ab}z_{0(cd)}
-\frac{2}{3}t^{ab}t_{a}{}^{c}z_{0b}{}^{d}z_{0dc}-\frac{13}{12}t^{ab}t_{a}{}^{c}z_{0b}{}^{d}z_{0cd}-\frac{7}{12}t^{ab}t_{a}{}^{c}z_{0}{}^{d}{}_{b}z_{0dc}+\nonumber\\
&+\frac{1}{4}t^{ab}z_{ab}{}^{c}t_{0}{}^{d}z_{0dc}+\frac{1}{4}t^{ab}t_{0}{}^{c}(z_{acd}z_{0b}{}^{d}-2z_{d(ac)}z_{0}{}^{d}{}_{b})\,.
\end{align}
\end{subequations}
These can be rewritten in terms of the Newton-Cartan curvatures described in \autoref{sec:Geometry} as
\begin{subequations}
\begin{align}
\accentset{(0)}{\mathcal{L}}_{\alpha}=&\,  \bigg[-\RicJ+2\CDRH_{a0}{}^{a}-\frac{3}{2}\RH_{0a}\RH_{0}{}^{a} \bigg]^{2}\,,\\
\nonumber\\
\accentset{(0)}{\mathcal{L}}_{\beta}=&\,  \RicJ_{ab}\RicJ^{ab}+2\RicJ_{ab}\CDRH^{ab}{}_{0}-2\RicJ_{0a}\CDRH^{ba}{}_{b}+\nonumber\\
&+2\RicJ_{0a}\RH^{ab}\RH_{0b}+\frac{1}{2}\RG_{0c}{}^{c}\RH^{ab}\RH_{ab}+\RicJ_{ab}\RH_{0}{}^{a}\RH_{0}{}^{b}+\nonumber\\
&-\CDRH_{a}{}^{ab}\CDRH_{00b}+\frac{1}{2}\CDRH_{a0}{}^{a}\CDRH_{b0}{}^{b}+\CDRH^{ab}{}_{0}\CDRH_{(ab)0}+\nonumber\\
&-\CDRH_{00a}\RH^{ab}\RH_{0b}-\CDRH_{a0}{}^{a}\RH_{0b}\RH_{0}{}^{b}+\CDRH^{ab}{}_{0}\RH_{0a}\RH_{0b}+\nonumber\\
&+\frac{3}{4} \RH_{0}{}^{a} \RH_{0a} \RH_{0}{}^{b} \RH_{0b}\,,\\
\nonumber\\
\accentset{(0)}{\mathcal{L}}_{\gamma}=&\,  \RJ^{abcd}\RJ_{abcd}-\frac{10}{3}\RJ_{0}{}^{(ab)c}\CDRH_{abc}+2\RG^{b(ac)}\CDRH_{abc}+\nonumber\\
&+2\RG_{0ab}\RH^{ac}\RH^{b}{}_{c}+\frac{8}{3}\RJ_{0}{}^{(ac)b}\RH_{ab}\RH_{0c}+\nonumber\\
&-2\CDRH_{00}{}^{a}\RH_{ab}\RH_{0}{}^{b}+2\CDRH^{ab}{}_{0}\RH_{0a}\RH_{0b}+\nonumber\\
&-2\CDRH_{0}{}^{ab}\CDRH_{0ab}+2\CDRH^{ab}{}_{0}\CDRH_{ab0}+\frac{3}{4}\RH_{0}{}^{a} \RH_{0 a} \RH_{0}{}^{b} \RH_{0 b}\,,
\end{align}
\end{subequations}
where $\CDRH_{\mu\nu\rho}=\nablaNR_{\mu}\RH_{\nu\rho}$. Our results show that, to take the non-relativistic limit of higher-derivatives gravity theories at the level of the action, the strategy \cite{Bergshoeff:2015uaa} of using a gauge field to cancel the divergent part can still be adopted with some  important remarks:
\begin{itemize}
\item in contrast with the two-derivative case, the full cancellation divergences for higher-derivative theories requires non-trivial higher-derivative couplings between the gravity and the gauge field sector,
\item it is not necessary to add more than a 1-form gauge field to achieve full cancellation of the divergences,
\item the cancellation terms do not depend on the spacetime dimension. 
\end{itemize}

Before studying the limit of the equations of motion we briefly comment on the higher-dimensional origin of the results that we have just obtained. 

\subsection{Non-Relativistic Limit and Higher-Dimensional Origin} \label{sec:HigherDimOrigin}
\noindent  Remarkably the relativistic actions in \autoref{eq:ActionsRel}, for which the non-relativistic magnetic limit is well defined, can be oxidized to a pure gravitational theory in one dimension higher.  In particular, these pure gravitational theories in $D+1$ dimensions correspond to the pure gravitational sector of each of the actions in  \autoref{eq:ActionsRel}. The theories appearing in \autoref{eq:ActionsRel} and their oxidation are linked via the following ansatz
\begin{align}
G_{MN}=&\,  
\renewcommand{\arraystretch}{1.8}
\begin{pNiceMatrix}[first-row, first-col,]
& \mu&x \\
\nu\,& g_{\mu\nu}+\Phi A_{\mu}A_{\nu} & \Phi A_{\nu} \\
x\,& \Phi A_{\mu} & \Phi \\
\end{pNiceMatrix}\,,& G^{MN}=&\,  
\renewcommand{\arraystretch}{1.8}
\begin{pNiceMatrix}[first-row, first-col,]
& \mu&x \\
\nu\,& g^{\mu\nu} & -A^{\nu} \\
x\,&-A^{\mu} & A^{\rho}A_{\rho}+\dfrac{1}{\Phi} \\
\end{pNiceMatrix}\,,
\end{align}
after setting the scalar field $\Phi$ to one; where $x$ is the isometry direction. This means that the coefficients of the different terms in $\mathcal{L}_{\CA}$, $\mathcal{L}_{\CB}$ and $\mathcal{L}_{\CC}$, which are fixed by the request of no divergences in the limit $\cc{}\rightarrow \infty$, are the same as those obtained by considering the pure gravitational part of these actions in one dimension higher and performing a Kaluza-Klein reduction on a circle. For example \autoref{eq:ActionsRelA} can be obtained from the $(D+1)$-dimensional Ricci scalar squared theory.

\begin{figure}[h!]
\centering 
\resizebox{\textwidth}{!}{

\tikzset{every picture/.style={line width=0.75pt}} 

\begin{tikzpicture}[x=0.75pt,y=0.75pt,yscale=-1,xscale=1]

\draw [color={rgb, 255:red, 0; green, 0; blue, 0 }  ,draw opacity=1 ][fill={rgb, 255:red, 128; green, 128; blue, 128 }  ,fill opacity=1 ][line width=1.5]    (410,170) -- (856.32,358.44) ;
\draw [shift={(860,360)}, rotate = 202.89] [fill={rgb, 255:red, 0; green, 0; blue, 0 }  ,fill opacity=1 ][line width=0.08]  [draw opacity=0] (13.4,-6.43) -- (0,0) -- (13.4,6.44) -- (8.9,0) -- cycle    ;
\draw [line width=1.5]    (411.5,170) -- (411.5,224.6) -- (411.5,334)(408.5,170) -- (408.5,224.6) -- (408.5,334) ;
\draw [shift={(410,344)}, rotate = 270] [fill={rgb, 255:red, 0; green, 0; blue, 0 }  ][line width=0.08]  [draw opacity=0] (13.4,-6.43) -- (0,0) -- (13.4,6.44) -- (8.9,0) -- cycle    ;
\draw [line width=2.25]    (500,390) .. controls (501.67,388.33) and (503.33,388.33) .. (505,390) .. controls (506.67,391.67) and (508.33,391.67) .. (510,390) .. controls (511.67,388.33) and (513.33,388.33) .. (515,390) .. controls (516.67,391.67) and (518.33,391.67) .. (520,390) .. controls (521.67,388.33) and (523.33,388.33) .. (525,390) .. controls (526.67,391.67) and (528.33,391.67) .. (530,390) .. controls (531.67,388.33) and (533.33,388.33) .. (535,390) .. controls (536.67,391.67) and (538.33,391.67) .. (540,390) .. controls (541.67,388.33) and (543.33,388.33) .. (545,390) .. controls (546.67,391.67) and (548.33,391.67) .. (550,390) .. controls (551.67,388.33) and (553.33,388.33) .. (555,390) .. controls (556.67,391.67) and (558.33,391.67) .. (560,390) .. controls (561.67,388.33) and (563.33,388.33) .. (565,390) .. controls (566.67,391.67) and (568.33,391.67) .. (570,390) .. controls (571.67,388.33) and (573.33,388.33) .. (575,390) .. controls (576.67,391.67) and (578.33,391.67) .. (580,390) .. controls (581.67,388.33) and (583.33,388.33) .. (585,390) .. controls (586.67,391.67) and (588.33,391.67) .. (590,390) .. controls (591.67,388.33) and (593.33,388.33) .. (595,390) .. controls (596.67,391.67) and (598.33,391.67) .. (600,390) .. controls (601.67,388.33) and (603.33,388.33) .. (605,390) .. controls (606.67,391.67) and (608.33,391.67) .. (610,390) .. controls (611.67,388.33) and (613.33,388.33) .. (615,390) .. controls (616.67,391.67) and (618.33,391.67) .. (620,390) .. controls (621.67,388.33) and (623.33,388.33) .. (625,390) .. controls (626.67,391.67) and (628.33,391.67) .. (630,390) .. controls (631.67,388.33) and (633.33,388.33) .. (635,390) .. controls (636.67,391.67) and (638.33,391.67) .. (640,390) .. controls (641.67,388.33) and (643.33,388.33) .. (645,390) .. controls (646.67,391.67) and (648.33,391.67) .. (650,390) .. controls (651.67,388.33) and (653.33,388.33) .. (655,390) .. controls (656.67,391.67) and (658.33,391.67) .. (660,390) .. controls (661.67,388.33) and (663.33,388.33) .. (665,390) .. controls (666.67,391.67) and (668.33,391.67) .. (670,390) .. controls (671.67,388.33) and (673.33,388.33) .. (675,390) .. controls (676.67,391.67) and (678.33,391.67) .. (680,390) .. controls (681.67,388.33) and (683.33,388.33) .. (685,390) .. controls (686.67,391.67) and (688.33,391.67) .. (690,390) .. controls (691.67,388.33) and (693.33,388.33) .. (695,390) .. controls (696.67,391.67) and (698.33,391.67) .. (700,390) .. controls (701.67,388.33) and (703.33,388.33) .. (705,390) .. controls (706.67,391.67) and (708.33,391.67) .. (710,390) .. controls (711.67,388.33) and (713.33,388.33) .. (715,390) .. controls (716.67,391.67) and (718.33,391.67) .. (720,390) .. controls (721.67,388.33) and (723.33,388.33) .. (725,390) .. controls (726.67,391.67) and (728.33,391.67) .. (730,390) .. controls (731.67,388.33) and (733.33,388.33) .. (735,390) -- (737,390) -- (745,390) ;
\draw [shift={(750,390)}, rotate = 180] [fill={rgb, 255:red, 0; green, 0; blue, 0 }  ][line width=0.08]  [draw opacity=0] (16.07,-7.72) -- (0,0) -- (16.07,7.72) -- (10.67,0) -- cycle    ;

\draw  [fill={rgb, 255:red, 80; green, 227; blue, 194 }  ,fill opacity=1 ]  (294,111) .. controls (294,108.24) and (296.24,106) .. (299,106) -- (527,106) .. controls (529.76,106) and (532,108.24) .. (532,111) -- (532,167) .. controls (532,169.76) and (529.76,172) .. (527,172) -- (299,172) .. controls (296.24,172) and (294,169.76) .. (294,167) -- cycle  ;
\draw (413,139) node  [font=\Large] [align=left] {\begin{minipage}[lt]{159.71pt}\setlength\topsep{0pt}
\begin{center}
$\displaystyle ( D+1)$-dim Relativistic \\Higher-Order Gravity
\end{center}

\end{minipage}};
\draw  [fill={rgb, 255:red, 72; green, 56; blue, 134 }  ,fill opacity=1 ]  (751,363) .. controls (751,360.24) and (753.24,358) .. (756,358) -- (967,358) .. controls (969.76,358) and (972,360.24) .. (972,363) -- (972,420) .. controls (972,422.76) and (969.76,425) .. (967,425) -- (756,425) .. controls (753.24,425) and (751,422.76) .. (751,420) -- cycle  ;
\draw (861.5,391.5) node  [font=\Large,color={rgb, 255:red, 255; green, 255; blue, 255 }  ,opacity=1 ] [align=left] {\begin{minipage}[lt]{147.82pt}\setlength\topsep{0pt}
\begin{center}
$\displaystyle D${\fontfamily{ptm}\selectfont -dim Non-Relativistic }\\{\fontfamily{ptm}\selectfont Higher-Order Theories}
\end{center}

\end{minipage}};
\draw (634.47,249.06) node  [font=\large,rotate=-23.29] [align=left] {{\footnotesize null-reduction}};
\draw (375.5,254.5) node  [font=\large,rotate=-270] [align=left] {\begin{minipage}[lt]{88.72pt}\setlength\topsep{0pt}
\begin{center}
{\footnotesize spatial dimensional }\\{\footnotesize reduction}
\end{center}

\end{minipage}};
\draw (625,390) node  [font=\large] [align=left] {{\footnotesize non-relativistic limit}\\\\{\footnotesize $\displaystyle \ \ \ \ \ \ \ \cc{}\ \rightarrow \infty $}};
\draw  [fill={rgb, 255:red, 173; green, 173; blue, 173 }  ,fill opacity=1 ]  (218,410) .. controls (218,407.24) and (220.24,405) .. (223,405) -- (403,405) .. controls (405.76,405) and (408,407.24) .. (408,410) -- (408,527) .. controls (408,529.76) and (405.76,532) .. (403,532) -- (223,532) .. controls (220.24,532) and (218,529.76) .. (218,527) -- cycle  ;
\draw (313,468.5) node  [font=\Large] [align=left] {Others \ \\$\displaystyle D$-dim Relativistic \ \\Higher-Order \\Gravity + Maxwell};
\draw  [fill={rgb, 255:red, 217; green, 219; blue, 241 }  ,fill opacity=1 ]  (318,350) .. controls (318,347.24) and (320.24,345) .. (323,345) -- (497,345) .. controls (499.76,345) and (502,347.24) .. (502,350) -- (502,436) .. controls (502,438.76) and (499.76,441) .. (497,441) -- (323,441) .. controls (320.24,441) and (318,438.76) .. (318,436) -- cycle  ;
\draw (410,393) node  [font=\Large] [align=left] {\begin{minipage}[lt]{122.57pt}\setlength\topsep{0pt}
\begin{center}
$\displaystyle D$-dim Relativistic \\Higher-Order \\Gravity + Maxwell
\end{center}

\end{minipage}};

\end{tikzpicture}
}
\vspace{0.4cm}
\caption{The figure shows that both the higher-order relativistic theory and its non-relativistic limit are linked to the same higher-dimensional origin. Given the higher-order $(D+1)$-dimensional theory, we can, on one hand, perform a Kaluza-Klein reduction to land on the very particular $D$-dimensional relativistic higher-order theory that has no $\cc{}$-divergences, and then take the non-relativistic limit. On the other hand, we can land right away into the higher-order non-relativistic theory by performing a null-reduction. }
\label{fig:diagram1}
\end{figure}
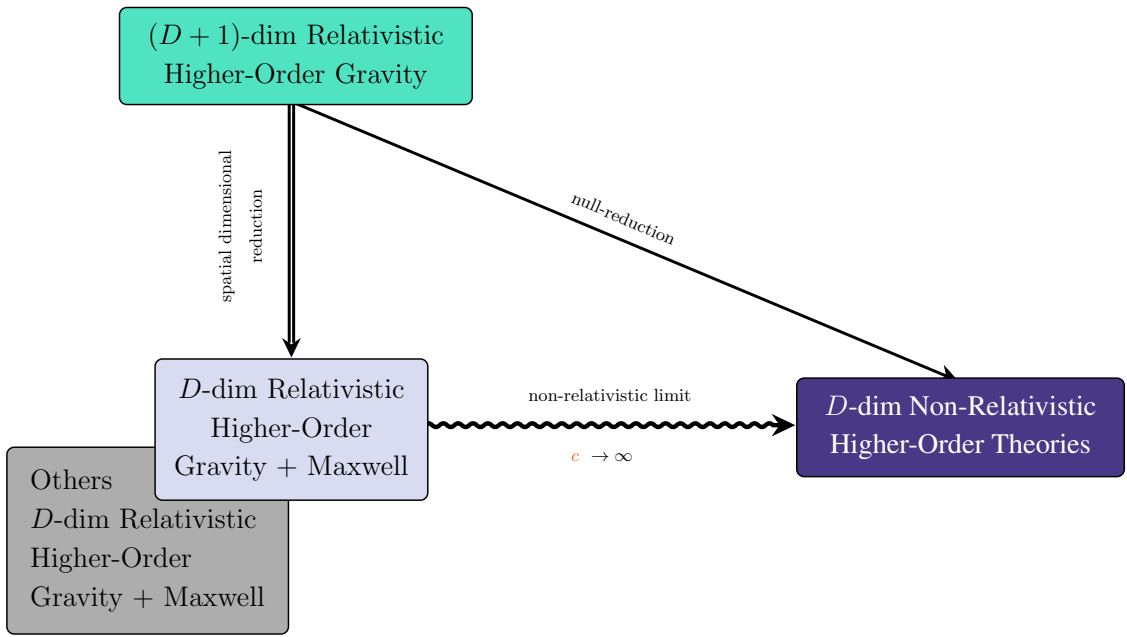

On top of this, the non-relativistic actions that we have obtained can also be obtained from the $(D+1)$-dimensional pure gravitational theory mentioned above via a null-reduction with the following ansatz
\begin{align}
G_{MN}=&\,   
\renewcommand{\arraystretch}{1.8}
\begin{pNiceMatrix}[first-row, first-col,]
& \mu&z \\
\nu\,& h_{\mu\nu}+2\tau_{(\mu}a_{\nu)} & -\tau_{\nu} \\
z\,& -\tau_{\mu} & 0 \\
\end{pNiceMatrix}\,,& G^{MN}=&\,  
\renewcommand{\arraystretch}{1.8}
\begin{pNiceMatrix}[first-row, first-col,]
& \mu&z \\
\nu\,& h^{\mu\nu} & h^{\nu\rho}a_{\rho}-\tau^{\nu} \\
z\,&h^{\mu\rho}a_{\rho}-\tau^{\mu} & h^{\rho\sigma}a_{\rho}a_{\sigma}-2\tau^{\rho}a_{\rho} \\
\end{pNiceMatrix}\,,
\end{align}
where the fields appearing in this ansatz are exactly those of Newton-Cartan geometry and $z$ is the null isometry direction. The scenario is displayed in \autoref{fig:diagram1}. Although we have considered the non-relativistic limit in a very specific set-up with a defined ansatz, our findings seem to strengthen the relation between relativistic theory with a null-isometry direction and non-relativistic theory, extending these bonds beyond the two-derivative case.

%
\FloatBarrier
\section{Non-Relativistic Limit of Equations of Motion \& the Poisson Equation}\label{sec:EomLimit}
\noindent  In the previous section we have studied the non-relativistic limit of the higher-derivative gravity action. In this section, we shift our attention to the limit taken at the level of the equations of motion, and, in particular, we investigate the possibility of obtaining the higher-derivative corrected version of the Poisson equation directly from this limit. We start by reviewing the two-derivative case and then we consider four-derivative corrections. 
\subsection{Einstein-Hilbert-Maxwell}\label{sec:GR}
\noindent  The Poisson equation is a second-order scalar differential equation describing, in Newtonian gravity, the relations between the Newton potential and the mass distribution.  It is well known that performing the non-relativistic limit at the level of the action does not guarantee that the resulting action contains the Poisson equation \cite{Bergshoeff:2024ipq}.
This happens in several cases \cite{Bergshoeff:2021tfn, Bergshoeff:2021bmc}. However, it is possible, in the same setting, to derive the Poisson equation from the limit of the equations of motion. The limit of the equations of motion is obtained by substituting the redefinition ansatz into them. This yields an expansion in powers of $\cc{}$, as 
\begin{align}
[X]=\cc{n}\accentset{(n)}{[X]}+\cc{n-2}\accentset{(n-2)}{[X]}+...=0\,,
\end{align}
where $[X]$ denotes the equation of motion of a generic field $X$  and $\accentset{(i)}{[X]}$ denotes the coefficient of the $i$-th power of $\cc{}$ in the expansion induced by the ansatz (we will adopt this notation for the rest of the paper).  Performing the limit $\cc{}\rightarrow \infty$ just amounts to selecting the leading order of its $\cc{}$-expansion. In this example thus the result gives:
\begin{align}
\accentset{(n)}{[X]}=&\,  0\,.
\end{align}
However, two or more equations of motion could have non-linearly independent leading terms in the expansion, implying that a proper combination of them will have an expansion starting at a lower power of $\cc{}$. This means that the result of the limit strongly depends on how the equations are combined before expanding and sending $\cc{}$ to infinity. The prescription that minimises the amount of terms lost in the limit is to combine them in such a way as to have the minimum number of powers of $\cc{}$ appearing in their expansion. For further details see \cite{Bergshoeff:2024ipq}. The general scalar combination that we can consider for the Poisson equation is:
\begin{align}
[P]:=&\,  \PA\, E^{\mu}{}_{a}E^{\nu a}[G]_{\mu\nu}+\PC E^{\mu}{}_{0}E^{\nu}{}_{0}\, [G]_{\mu\nu}+\PB\, E_{\mu}{}^{0}[A]^{\mu}\,,\label{eq:PoissonCombination}
\end{align}
where $\PA,\PB$ and $\PC$ are three constant parameters that we will try to fine-tune to get the Poisson equation from the limit and $[G]_{\mu\nu}$ and $[A]^{\mu}$ are the equations of motion for the metric and the gauge field, defined by
\begin{align}
\delta \mathcal{S}=&\,  \int\,d^{D}x\, \sqrt{-g} \Big([G]_{\mu\nu}\delta g^{\mu\nu}+[A]^{\mu}\delta A_{\mu}\Big)
\end{align}
and explicitly given, in the Einstein-Hilbert-Maxwell two-derivative case, by
\begin{subequations}
\begin{align}
[A]^{\mu}:=&\,  \nabla_{\nu}F^{\nu\mu}=0\,,\\
\nonumber\\
[G]_{\mu\nu}:=&\,  \Ric_{\mu\nu}-\frac{1}{4}g_{\mu\nu}\R-\frac{1}{2}F_{\mu\rho}F_{\nu}{}^{\rho}+\frac{1}{8}g_{\mu\nu}F^{2}=0\,.
\end{align}
\end{subequations}
Let us have a look at the expansion of $[P]$. We have 
\begin{align}
[P]=&\,  \cc{2}\accentset{(2)}{[P]}+\accentset{(0)}{[P]}+\cc{-2}\accentset{(-2)}{[P]}+\mathcal{O}(\cc{-2})\,,
\end{align}
with
\begin{subequations}
\begin{align}
\accentset{(2)}{[P]}=&\,  \frac{1}{4}(\PC-2\PB)t_{ab}t^{ab}\,,\label{eq:PtwoderexpansionOrder2}\\
\nonumber\\
\accentset{(0)}{[P]}=&\,  \frac{1}{2}(3\PA-D\PA+\PC)\RicNR_{a}{}^{a}+
(2\PA-D\PA+\PB)\Ct_{a0}{}^{a}+\nonumber\\
&+\frac{1}{4}(-5\PA-\PC+3D\PA)t_{0a}t_{0}{}^{a}+\frac{1}{2}(-3\PA-2\PB+D\PA)t^{ab}f_{ab}\,,\label{eq:PtwoderexpansionOrder0}\\ 
\nonumber\\
\accentset{(-2)}{[P]}=&\,  \frac{1}{2}(\PA-D\PA+2\PB-\PC)\Cm_{a0}{}^{a}+\frac{1}{2}(D\PA-\PA+\PC)\RicNR_{00}+\frac{1}{2}(3\PA+\PC-D\PA)\Cz_{00a}{}^{a}+\nonumber\\
&+\frac{1}{2}(3\PA-D\PA+\PC)\bigg(z_{0a}{}^{a}z_{0b}{}^{b}+z_{0}{}^{ab}z_{0[ab]}-\frac{1}{2}z_{0ab}f^{ab}-z^{ab}{}_{a}f_{0b}\bigg)+\nonumber\\
&+\frac{1}{4}(D\PA-3\PA-2\PB)f^{ab}f_{ab}+At_{0}{}^{a}f_{0a}\,.\label{eq:PtwoderexpansionOrderm2}
\end{align}\label{eq:Ptwoderexpansion}
\end{subequations}
The Poisson equation occurs at order $\cc{-2}$ since the two-derivative term acting on the Newton potential, identifiable with $\tau^{\mu}a_{\mu}$, 
\begin{align}
h^{\nu\rho}\partial_{\mu}\partial_{\nu}(\tau^{\mu}a_{\rho})
\end{align}
is contained in the term $\Cm_{a0}{}^{a}$ and in $\RicNR_{00}$. 
This means that to obtain the Poisson equation from the non-relativistic limit of the combination [P], we should have the vanishing of the highest orders $\cc{2}$ and $\cc{0}$. While for the term at order $\cc{2}$ this is possible just setting $\PC=2\PB$, for the term at order $\cc{0}$ a full cancellation cannot be achieved for any non-trivial choice of $\PA$ and $\PB$. This depends on the fact that, in the present case, we do not have an emergent dilatation symmetry. The presence of that symmetry would have helped us in achieving our goal as described in \cite{Bergshoeff:2024ipq}. Nevertheless, the absence of such a symmetry does not completely preclude obtaining the Poisson equation via the limit; the problem, indeed, could be circumvented by choosing opportunely $\PA, \PB$ and $\PC$ and imposing concomitantly a suitable constraint. One possibility is to impose \cite{Bergshoeff:2015uaa}
\begin{align}
F_{\mu\nu}=&\,  0\,.\label{eq:constr1}
\end{align}
This constraint is fully invariant under the set of relativistic transformations, implying that it does not require further constraints. By plugging in the limit ansatze \autoref{eq:ansatze}, it can also be rewritten as:
\begin{align}
t_{\mu\nu}=&\,  -\frac{1}{\cc{2}}f_{\mu\nu}.\label{eq:constraint1}
\end{align} 
Using these relations before taking the limit of \autoref{eq:Ptwoderexpansion} allows us to reduce the power of the terms proportional to the intrinsic torsion, $t_{\mu\nu}$. Thus the order $\cc{2}$ gets shifted to order $\cc{-2}$. The same applies to three out of the four terms appearing in $\accentset{(0)}{[P]}$. The fourth term can be canceled by setting $\PC=(D-3)\PA$ and then the leading power in the expansion will be the following $\cc{-2}$ term:
\begin{align}
\accentset{(-2)}{[P]}=&\,  (D-2)\PA\RicNR_{00}=(D-2)\PA\RG_{0a}{}^{a}\,,
\end{align}
where the different contributions appearing and disappearing with respect to \autoref{eq:PtwoderexpansionOrderm2} are due to those coming from \autoref{eq:PtwoderexpansionOrder0} and \autoref{eq:PtwoderexpansionOrder2} via \autoref{eq:constraint1}. Note that having used \autoref{eq:constraint1}  in \autoref{eq:Ptwoderexpansion} the torsion term in the connection is no longer present (with a slight abuse of notation, we have used the same symbols).\footnote{Before taking the non-relativistic limit and without using the constraint \autoref{eq:constr1} it would have been:\\
$\RicNR_{00}=\RG_{0a}{}^{a}-\tau^{\mu}\oJ_{\mu}{}^{a}\RH_{0a}$.} In the limit, the constraint \autoref{eq:constraint1} becomes the zero-torsion condition
\begin{align}
t_{\mu\nu}=0\,.
\end{align} 
It is not surprising that eventually $\PB$ does not appear since the constraint $F_{\mu\nu}=0$ implies $[A]^{\mu}=0$ automatically, thus dropping the corresponding term in \autoref{eq:PoissonCombination}.
 
To obtain the corrections to the Poisson equation we should replicate for the higher-order theories what we have done in this section. In doing so there are some relevant aspects to mention: 
\begin{itemize}
    \item since the two-derivative contribution is unmodified in the four-derivative theories, the Poisson equation will still occur at order $\cc{-2}$ in the expansion of \autoref{eq:PoissonCombination}.
 \item The constraint \autoref{eq:constraint1} should still be imposed to guarantee the two-derivative part to work. 
\end{itemize}
This means that, if the higher-derivative terms produce higher-order terms in the expansion \autoref{eq:Ptwoderexpansion}, such as order $\cc{4}$ or higher, we should make them vanish with a combination of constraints and fine-tuning of $\PA,\PB$ and $\PC$, to make order $\cc{-2}$ the leading one. The goal of the next subsection is to inspect the mechanism to get the Poisson equation with higher-derivative corrections, performing the non-relativistic limit, and disclose it in some particular cases.  We start by considering the Gauss-Bonnet gravity. 

\subsection{Einstein-Gauss-Bonnet Gravity}\label{sec:GBcase}
\noindent In between the two-derivative case and the four-derivative theory, there exists one particular case deserving a specific analysis, this is Einstein-Gauss-Bonnet gravity. This is the theory described by the following four-derivatives action: 
\begin{align}
\mathcal{S}_{GB}=&\,  \int d^{D}x\, \sqrt{-g}\Big[\R+\CA(\R^2-4\Ric_{\mu\nu}\Ric^{\mu\nu}+\Riem_{\mu\nu\rho\sigma}\Riem^{\mu\nu\rho\sigma})\Big]\,.\label{eq:GB3}
\end{align}
In four dimensions, the quadratic term is a topological invariant. Notice that, despite having an action that is quartic in the derivatives of the metric, the equations of motion are second order. For this reason this could be the perfect bridge between the two-derivative and four-derivative theories. Following \autoref{sec:NRActionLimit}, to ensure the finiteness of the limit at the level of the action, the Lagrangian in \autoref{eq:GB3} should be supplemented with the Maxwell terms. This leads to:
\begin{align}
\mathcal{S}_{GB}=&\,  \int d^{D}x\, \sqrt{-g}\Big[\mathcal{L}_{EHM}+\CA(\mathcal{L}_{\CA}-4\mathcal{L}_{\CB}+\mathcal{L}_{\CC})\Big]\,.
\end{align}
For this action the equations of motion are
\begin{subequations}
\begin{align}
[A]^{\nu}:=&\,  \nabla_{\mu}\bigg[F^{\mu\nu}-\frac{3}{2}\CA F^{2}F^{\mu\nu}+3\CA F^{\mu\alpha}F_{\alpha\beta}F^{\nu\beta}\bigg]+\nonumber\\
&+8\CA\Ric^{\rho[\mu}\nabla_{\mu}F^{\nu]}{}_{\rho}+\CA\Riem^{\mu\nu\rho\sigma}\nabla_{\mu}F_{\rho\sigma}+2\CA\nabla_{\mu}F^{\mu\nu}\R\,,\\
\nonumber\\
[G]_{\alpha\beta}:=&\,  \Ric_{\alpha\beta}-\frac{1}{2}g_{\alpha\beta}\R-\frac{1}{2}F_{\alpha\mu}F_{\beta}{}^{\mu}+\frac{1}{8}g_{\alpha\beta}F^{2}+\nonumber\\
&+\CA\bigg[-\frac{3}{2}F_{\alpha}{}^{\mu}F_{\mu}{}^{\nu}F_{\nu}{}^{\rho}F_{\rho\beta}+\frac{3}{4}F_{\alpha}{}^{\mu}F_{\beta\mu}F^{2}+\frac{3}{16}g_{\alpha\beta}F_{\sigma}{}^{\mu}F_{\mu}{}^{\nu}F_{\nu}{}^{\rho}F_{\rho}{}^{\sigma}-\frac{3}{32}g_{\alpha\beta}F^{4}+\nonumber\\
&+\frac{3}{4}g_{\alpha\beta}F^{\mu\nu}F^{\rho\sigma}\Riem_{\mu\nu\rho\sigma}+2F_{\mu}{}^{\nu}F^{\mu\rho}\Riem_{\alpha\rho\beta\nu}-\frac{3}{2}F^{\nu\rho}(F_{\beta}{}^{\mu}\Riem_{\alpha\mu\nu\rho}+F_{\alpha}{}^{\mu}\Riem_{\beta\mu\nu\rho})+\nonumber\\
&-\frac{1}{2}\Ric_{\alpha\beta}F^{2}-2F_{\beta}{}^{\mu}F_{\mu}{}^{\nu}\Ric_{\alpha\nu}-2F_{\alpha}{}^{\mu}F_{\mu}{}^{\nu}\Ric_{\beta\nu}+3F_{\alpha}{}^{\mu}F_{\beta}{}^{\nu}\Ric_{\mu\nu}+\nonumber\\
&-2g_{\alpha\beta}F_{\mu}{}^{\nu}F^{\mu\rho}\Ric_{\nu\rho}-F_{\alpha\mu}F_{\beta}{}^{\mu}\R+\frac{1}{4}g_{\alpha\beta}F^{2}\R+\nonumber\\
&+\frac{1}{2}\nabla_{\alpha}F_{\mu\nu}\nabla_{\beta}F^{\mu\nu}
-\nabla_{\mu}F_{\alpha}{}^{\mu}\nabla_{\nu}F_{\beta}{}^{\nu}
+\nabla_{\alpha}F_{\beta\mu}\nabla_{\nu}F^{\mu\nu}
+\nabla_{\beta}F_{\alpha\mu}\nabla_{\nu}F^{\mu\nu}
+\nabla_{\mu}F_{\alpha\nu}\nabla^{\mu}F_{\beta}{}^{\nu}+\nonumber\\
&-\frac{1}{2}g_{\alpha\beta}\nabla_{\mu}F_{\nu\rho}\nabla^{\mu}F^{\nu\rho}-g_{\alpha\beta}\nabla_{\mu}F^{\mu\nu}\nabla^{\rho}F_{\nu\rho}+\nonumber\\
&-4\Ric_{\alpha}{}^{\mu}\Ric_{\beta\mu}+2g_{\alpha\beta}\Ric_{\mu\nu}\Ric^{\mu\nu}+2\Ric_{\alpha\beta}\R -\frac{1}{2}g_{\alpha\beta}\R^{2}+\nonumber\\
&-4\Ric^{\mu\nu}\Riem_{\alpha\mu\beta\nu}+2\Riem_{\alpha}{}^{\mu\nu\rho}\Riem_{\beta\mu\nu\rho}-\frac{1}{2}g_{\alpha\beta}\Riem_{\mu\nu\rho\sigma}\Riem^{\mu\nu\rho\sigma}\bigg]\,,
\end{align}
\end{subequations}
where they have been rewritten in the best way to study their expansion using the relations listed in \autoref{sec:UsRel}. Then for the Poisson combination \autoref{eq:PoissonCombination} we have:
\begin{align}
[P]=&\,  \cc{2}\accentset{(2)}{[P]}+\accentset{(0)}{[P]}+\cc{-2}\accentset{(-2)}{[P]}+\mathcal{O}(\cc{-2})\,,
\end{align}
with
\begin{subequations}
\begin{align}
\accentset{(2)}{[P]}=&\,  (\PC-2\PB)\CA\bigg[\frac{1}{4}t_{ab}t^{ab}\bigg(\CA^{-1}+2\RicNR_{c}{}^{c}+\frac{1}{3}t_{0}{}^{c}t_{0c}-2t^{cd}f_{cd}\bigg)+\nonumber\\
&+t^{ab}\bigg(-t_{0a}\Ct_{cb}{}^{c}+\frac{2}{3}t_{0}{}^{c}\Ct_{(ac)b}-t_{a}{}^{c}\RicNR_{bc}-\frac{1}{6}t_{a}{}^{c}t_{0b}t_{0c}+t_{a}{}^{c}t_{b}{}^{d}f_{cd}\bigg)+\nonumber\\
&+\Ct_{a}{}^{ab}\Ct_{cb}{}^{c}+\frac{2}{3}\Ct^{abc}\Ct_{(ab)c}\bigg]
\,,\\
\nonumber\\
\accentset{(0)}{[P]}=&\,  \frac{1}{2}(3\PA+\PC-D\PA)\RicNR_{a}{}^{a}+\frac{1}{2}(5\PA+\PC-D\PA)\CA \Big[(\RicNR_{a}{}^{a})^{2}-4\RicNR^{ab}\RicNR_{ab}+\RiemNR^{abcd}\RiemNR_{abcd}\Big]+...\,, \label{eq:PoissonGBExpansion0}
\end{align}
\label{eq:PoissonGBExpansion}
\end{subequations}
where dots denote terms proportional to the intrinsic torsion. We realize that at order $\cc{2}$ we have all terms at least quadratic in the intrinsic torsion. This implies that the constraint \autoref{eq:constr1}
is able to shift completely the term appearing at order $\cc{2}$ to order $\cc{-2}$ or lower.  For the term appearing at order $\cc{0}$ in \autoref{eq:PoissonGBExpansion} all the terms proportional to the intrinsic torsion, not explicitly shown, are shifted to order $\cc{-2}$ or lower by the constraint. Therefore we are left with the terms appearing explicitly in \autoref{eq:PoissonGBExpansion0}.  Those terms have two different coefficients; thus, the fine-tuning of $\PC$ defined in \autoref{sec:GR} is not sufficient to cancel both at the same time. This prevents us from getting the Poisson equation from the limit. To overcome this issue, we could impose a further constraint, but we prefer to propose the following trick. 

\subsubsection{Scalar Field Trick}\label{sec:SclarFieldTrick}
\noindent Given a relativistic Lagrangian $\mathcal{L}=\mathcal{L}_{1}+\mathcal{L}_{2}$, we would like to extract the Poisson equation by taking the non-relativistic limit of its equations of motion. We notice that it, in certain cases, it is helpful to introduce a scalar field $\Phi$ and define a new Lagrangian, as
\begin{align}
\mathcal{L}_{\text{new}}=\mathcal{L}_{1}+{\rm e}^{\BA \Phi}\Big[\mathcal{L}_{2}+\BB \partial_{\mu}\Phi\partial^{\mu}\Phi\Big]\label{eq:DilatonTrick1}
\end{align}
with $\BA$ and $\BB$ two constants. Then the field equations become:
\begin{subequations}
\begin{align}
[\Phi]^{\text{new}}:=&\,  \BA \mathcal{L}_{2}+ \partial_{\mu}\Phi(...)^{\mu}\,,\\
[G]^{\text{new}}_{\mu\nu}:=&\,  [G]^{(1)}_{\mu\nu}+{\rm e}^{\BA \Phi}[G]^{(2)}_{\mu\nu}+\partial_{\mu}\Phi(...)^{\mu}\,,\\
[A]^{\text{new}}_{\mu}:=&\,  [A]^{(1)}_{\mu}+{\rm e}^{\BA \Phi}[A]^{(2)}_{\mu}+\partial_{\mu}\Phi(...)^{\mu}\,,
\end{align}\label{eq:EOMnew}
\end{subequations} 
where the superscripts $^{(1)}$ and $^{(2)}$ denote respectively the contributions to the field equation coming from $\mathcal{L}_{1}$ and $\mathcal{L}_{2}$ and dots denote irrelevant contribution for our treatment. Now consider the trivial ansatz for the scalar field:
\begin{align}
\Phi=&\,  \phi\,.
\end{align}
Imposing the constraint
\begin{align}
\partial_{\mu}\Phi=&\,  0\,,
\end{align}
implies that \autoref{eq:EOMnew} become
\begin{subequations}
\begin{align}
[\phi]^{\text{new}}=&\,  \BA \mathcal{L}_{2}\,,\\
[G]^{\text{new}}_{\mu\nu}=&\,  [G]^{(1)}_{\mu\nu}+{\rm e}^{\BA \phi}[G]^{(2)}_{\mu\nu}\,,\\
[A]^{\text{new}}_{\mu}=&\,  [A]^{(1)}_{\mu}+{\rm e}^{\BA \Phi}[A]^{(2)}_{\mu}\,.
\end{align}
\end{subequations}
The equation of motion of the gauge field does not play any role as it identically vanishes by the constraint $F_{\mu\nu}=0$. The deformation that we have introduced in our original model allows us to have an extra equation to try to cancel the leading and sub-leading terms of the expansion of the Poisson combination.  We can see how this works in the case of interest. We choose $\mathcal{L}_{2}$ to be the Einstein-Hilbert-Maxwell term and $\mathcal{L}_{1}$ the Gauss-Bonnet term. We recall that the Gauss-Bonnet Lagrangian at order zero expands as
\begin{align}
\mathcal{L}_{GB}=&\,  \RicNR_{a}{}^{a}+(\RicNR_{a}{}^{a})^{2}-4\RicNR_{ab}\RicNR^{ab}+\RiemNR^{abcd}\RiemNR_{abcd}+...\,,
\end{align}
where the dots denote terms proportional to the intrinsic torsion, in such a way that the constraint $F_{\mu\nu}=0$ could shift them to lower order if this term appears in the equations of motion.  Considering the combination
\begin{align}
[P]:=&\,  \PA\, E^{\mu}{}_{a}E^{\nu a}[G]_{\mu\nu}+\PC E^{\mu}{}_{0}E^{\nu}{}_{0}\, [G]_{\mu\nu}-\PB\, E_{\mu 0}[A]^{\mu}+\PD [\phi]\,,\label{eq:PoissonCombinationRevised}
\end{align}
the expansion is unmodified at order $\cc{2}$, while, at order $\cc{0}$, we get
\begin{align}
\accentset{(0)}{[P]}=&\,  \frac{1}{2}(3\PA+\PC-D\PA+2\PD\BA)\RicNR_{a}{}^{a}+\nonumber\\
&+\frac{1}{2}(5\PA+\PC-D\PA)\Big[(\RicNR_{a}{}^{a})^{2}-4\RicNR^{ab}\RicNR_{ab}+\RiemNR^{abcd}\RiemNR_{abcd}\Big]\,,
\end{align}
where we have used the constraints $F_{\mu\nu}=0$ to shift the terms proportional to the intrinsic torsion to a lower order in $\cc{}$. We have used the relations \autoref{sec:NRCurvaturesIds} and for simplicity, without losing generality, we have set $\phi=0$ (a different constant value amounts to a relative constant between the two- and four-derivative terms that can be reabsorbed in a redefinition). This implies that the cancellation can be realized by taking $\PC$ and $\PD$ to be
\begin{align}
C=&\,  (D-5)\PA\,,&\PD\BA=&\,  \PA\,.
\end{align}
We have been able to cancel the order $\cc{2}$ and $\cc{0}$ in the Poisson combination. The limit gives the following result for the Poisson equation:
\begin{align}
\accentset{(-2)}{[P]}=&\,   (D-4)\PA\RG_{0a}{}^{a} +\CA(D-4)\PA\bigg[4\RG_{ab}{}^{a}\RG^{bc}{}_{c}+\frac{1}{2}\RG^{abc}\RG_{abc}+\nonumber\\
&-2\RG_{0a}{}^{a}\RicJ_{b}{}^{b}+4\RG_{0}{}^{ab}\RicJ_{ab}+\frac{1}{2}\RJ^{0abc}(\RJ_{0abc}-2\RG_{bca})\bigg]\,.
\end{align}
It is not surprising that for $D=4$ the Poisson equation trivialises, since the Gauss-Bonnet term is topological in this number of dimensions.  The fact that the first term vanishes is because we have already fine-tuned the coupling and constants using the information coming from the higher-order terms; thus, in the absence of these higher-derivative corrections, such analysis should be redone \cite{Bergshoeff:2024ipq}. This example extends and enforces the results of \cite{Bergshoeff:2024ipq} where it was first shown the crucial role that the scalar field plays in getting the Poisson equation. The same results would have been obtained without inserting a scalar field, but imposing a second on-shell constraint, namely
\begin{align}
\R^{2}-4\Ric_{\mu\nu}\Ric^{\mu\nu}+\Riem_{\mu\nu\rho\sigma}\Riem^{\mu\nu\rho\sigma}=&\,  0\,.
\end{align}

\subsubsection{Full Set of Equations of Motions}\label{sec:NRGBEOM}
\noindent We now present the full set of non-relativistic equations of motion and constraints. The two constraints that we have imposed are the following:
\begin{subequations}
\begin{align}
\RH_{\mu\nu}=&\,  0\,,\\
\phi=&\,  0\,,
\end{align}
while the equations of motion are 
\begin{align}
[\phi]:=&\,  \RicJ=0\,,\\
\nonumber\\
\accentset{(-2)}{[P_{+}]}:=&\,  \RG_{0a}{}^{a} +4\RG_{ab}{}^{a}\RG^{bc}{}_{c}+\frac{1}{2}\RG^{abc}\RG_{abc}-2\RG_{0a}{}^{a}\RicJ+\nonumber\\
&+4\RG_{0}{}^{ab}\RicJ_{ab}+\frac{1}{2}\RJ^{0abc}(\RJ_{0abc}-2\RG_{bca})=0\,,\\
\nonumber\\
\accentset{(0)}{[P_{-}]}:=&\,  \RJ^{abcd}\RJ_{abcd}-4\RicJ^{ab}\RicJ_{ab}=0\,,\\
\nonumber\\
\accentset{(-1)}{[G]}_{0a}:=&\,  \RG_{ab}{}^{b}-2\RJ^{bc}{}_{a}{}^{d}\RG_{bcd}+4\RicJ_{ab}
\RG^{bc}{}_{c}-2\RicJ\RG_{ac}{}^{c}+\nonumber\\
&+4\RicJ_{ab}\RG^{bc}{}_{c}+4\RicJ^{bc}\RG_{abc}=0\,,\\
\nonumber\\
\accentset{(0)}{[G]}_{\{ab\}}:=&\,  -\RicJ_{ab}+2\RJ_{a}{}^{cde}\RJ_{bcde}-4\RicJ_{a}{}^{c}\RicJ_{bc}-4\RicJ^{cd}\RJ_{acbd}=0\,,
\end{align}
\end{subequations}
where, for simplicity, we have omitted the coupling $\alpha$ and rescaled the equations by overall factors. $\accentset{(-2)}{[P_{+}]}$ is the Poisson equation studied in the previous paragraph and $\accentset{(0)}{[P_{-}]}$ is obtained by taking the limit of the combination of the relativistic equations of motion  (up to an overall rescaling)  
\begin{align*}
[P_{-}]:=E^{\mu}{}_{a}E^{\nu a}[G]_{\mu\nu}+(D-3) E^{\mu}{}_{0}E^{\nu}{}_{0}\, [G]_{\mu\nu}-[\phi]^{2}\,,
\end{align*}
being the leading order at $\cc{0}$. $\accentset{(-1)}{[G]}_{0a}$ and $\accentset{(0)}{[G]}_{\{ab\}}$ are obtained by taking the limit of 
\begin{subequations}
\begin{align}
[G]_{0a}=&\,  E^{\mu}{}_{0}E^{\mu}_{a}[G]_{\mu\nu}\,,\\
[G]_{\{ab\}}=&\,  E^{\mu}{}_{\{a}E^{\nu}{}_{b\}}[G]_{\mu\nu}\,,
\end{align}\label{eq:g0agab}
\end{subequations}
where the curly brackets denote the traceless symmetric part, defined by
\begin{align*}
 T_{\{ab\}}=&\,  \frac{1}{2}\bigg(T_{ab}+T_{ba}-\frac{2}{D-1}\delta_{ab}T_{c}{}^{c}\bigg)\,.
 \end{align*}
The equations \autoref{eq:g0agab} have leading orders $\cc{-1}$ and $\cc{0}$ respectively. All the equations, except the Poisson, can be obtained from the corresponding non-relativistic action built as described in \autoref{sec:NRActionLimit}.

\begin{figure}[h!]
\centering
\resizebox{0.35\textwidth}{!}{%

\tikzset{every picture/.style={line width=0.75pt}} 

\begin{tikzpicture}[x=0.75pt,y=0.75pt,yscale=-1,xscale=1]

\draw  [fill={rgb, 255:red, 249; green, 249; blue, 237 }  ,fill opacity=1 ]  (695.5,71) .. controls (695.5,69.34) and (696.84,68) .. (698.5,68) -- (764.5,68) .. controls (766.16,68) and (767.5,69.34) .. (767.5,71) -- (767.5,140) .. controls (767.5,141.66) and (766.16,143) .. (764.5,143) -- (698.5,143) .. controls (696.84,143) and (695.5,141.66) .. (695.5,140) -- cycle  ;
\draw (731.5,105.5) node  [font=\huge]  {$\accentset{( -2)}{[ P_{+}]}$};
\draw  [fill={rgb, 255:red, 217; green, 219; blue, 241 }  ,fill opacity=1 ]  (690,237) .. controls (690,235.34) and (691.34,234) .. (693,234) -- (770,234) .. controls (771.66,234) and (773,235.34) .. (773,237) -- (773,314) .. controls (773,315.66) and (771.66,317) .. (770,317) -- (693,317) .. controls (691.34,317) and (690,315.66) .. (690,314) -- cycle  ;
\draw (731.5,275.5) node  [font=\huge]  {$\accentset{( -1)}{[ G]}_{0a}$};
\draw  [fill={rgb, 255:red, 182; green, 226; blue, 211 }  ,fill opacity=1 ]  (821,411) .. controls (821,409.34) and (822.34,408) .. (824,408) -- (889,408) .. controls (890.66,408) and (892,409.34) .. (892,411) -- (892,480) .. controls (892,481.66) and (890.66,483) .. (889,483) -- (824,483) .. controls (822.34,483) and (821,481.66) .. (821,480) -- cycle  ;
\draw (856.5,445.5) node  [font=\huge]  {$\accentset{( 0)}{[ P_{-}]}$};
\draw  [fill={rgb, 255:red, 184; green, 233; blue, 134 }  ,fill opacity=0.74 ]  (556,407) .. controls (556,405.34) and (557.34,404) .. (559,404) -- (654,404) .. controls (655.66,404) and (657,405.34) .. (657,407) -- (657,484) .. controls (657,485.66) and (655.66,487) .. (654,487) -- (559,487) .. controls (557.34,487) and (556,485.66) .. (556,484) -- cycle  ;
\draw (606.5,445.5) node  [font=\huge]  {$\accentset{( 0)}{[ G]}_{\{ab\}}$};
\draw  [fill={rgb, 255:red, 190; green, 190; blue, 190 }  ,fill opacity=1 ]  (703,413) .. controls (703,411.34) and (704.34,410) .. (706,410) -- (757,410) .. controls (758.66,410) and (760,411.34) .. (760,413) -- (760,478) .. controls (760,479.66) and (758.66,481) .. (757,481) -- (706,481) .. controls (704.34,481) and (703,479.66) .. (703,478) -- cycle  ;
\draw (731.5,445.5) node  [font=\huge]  {$\accentset{( 0)}{[ \Phi ]}$};
\draw [line width=1.5]    (731.5,143) -- (731.5,230) ;
\draw [shift={(731.5,234)}, rotate = 270] [fill={rgb, 255:red, 0; green, 0; blue, 0 }  ][line width=0.08]  [draw opacity=0] (20.36,-9.78) -- (0,0) -- (20.36,9.78) -- (13.52,0) -- cycle    ;
\draw [line width=1.5]    (700.99,317) -- (639.38,400.78) ;
\draw [shift={(637.01,404)}, rotate = 306.33] [fill={rgb, 255:red, 0; green, 0; blue, 0 }  ][line width=0.08]  [draw opacity=0] (20.36,-9.78) -- (0,0) -- (20.36,9.78) -- (13.52,0) -- cycle    ;
\draw [line width=1.5]    (762.01,317) -- (826.56,404.78) ;
\draw [shift={(828.93,408)}, rotate = 233.67] [fill={rgb, 255:red, 0; green, 0; blue, 0 }  ][line width=0.08]  [draw opacity=0] (20.36,-9.78) -- (0,0) -- (20.36,9.78) -- (13.52,0) -- cycle    ;
\draw [line width=1.5]    (731.5,317) -- (731.5,406) ;
\draw [shift={(731.5,410)}, rotate = 270] [fill={rgb, 255:red, 0; green, 0; blue, 0 }  ][line width=0.08]  [draw opacity=0] (20.36,-9.78) -- (0,0) -- (20.36,9.78) -- (13.52,0) -- cycle    ;

\end{tikzpicture}
}
\vspace{0.5cm}
\caption{We show the multiplet structure of the non-relativistic equations of motion of Gauss-Bonnet-Newton-Cartan gravity and how they transform under boost. The number in parentheses over the equations, following our conventions, indicates the power of $\cc{}$ at which the equation occurs when taking the limit of the relativistic equation. Each arrow points toward the equation of motion appearing in the transformation under a Galilean boost of the equation from which it departs. The set of equations of motion is organised in a reducible, indecomposable representation of the Bargmann algebra.}
\label{fig:diagram2}
\end{figure}
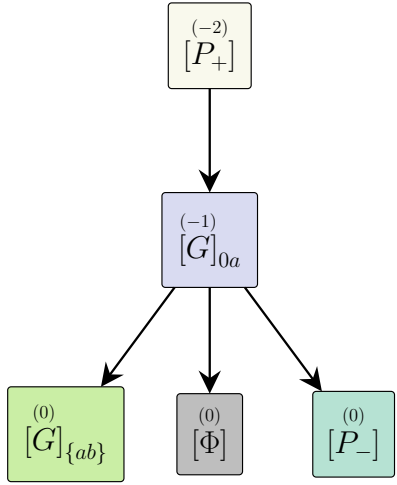

It is useful to report here the transformations under boost of the equations of motion and constraints
\begin{subequations}
\begin{align}
\delta_{G}\phi=&\,   0\,,\\
\delta_{G}\RH_{\mu\nu}=&\,   0\,,\\
\delta_{G}[\phi]=&\,   0\,,\\
\delta_{G}\accentset{(-2)}{[P_{+}]}=&\,  -2\boostp^{a}\accentset{(-1)}{[G]}_{0a}\,,\\
\delta_{G}\accentset{(0)}{[P_{-}]}=&\,   0\,,\\
\delta_{G}\accentset{(-1)}{[G]}_{0a}=&\,  -\boostp^{b}\accentset{(0)}{[G]}_{\{ab\}}-2\boostp_{a}\accentset{(0)}{[P_{-}]}+\boostp_{a}[\phi](1-2[\phi])\,,\\
\delta_{G}\accentset{(0)}{[G]}_{\{ab\}}=&\,   0\,,
\end{align}
\end{subequations}
where the right-hand side is up to the constraints. This shows that the non-relativistic equations of motion obtained by taking the limit of Gauss-Bonnet gravity with higher-derivative Maxwell forms a closed set of non-trivial equations. These describe zero-torsion Gauss-Bonnet Newton-Cartan gravity. The multiplet structure under boosts is depicted in \autoref{fig:diagram2}.
\FloatBarrier
\subsection{Quadratic Ricci Scalar Gravity}\label{sec:HigherOrderEOM}
\begin{figure}[t]
\centering
\resizebox{0.35\textwidth}{!}{%
\tikzset{every picture/.style={line width=0.75pt}} 
\begin{tikzpicture}[x=0.75pt,y=0.75pt,yscale=-1,xscale=1]
The full set of equations of motion and constraints defines quadratic Ricci scalar Newton-Cartan gravity.

\draw  [fill={rgb, 255:red, 249; green, 249; blue, 237 }  ,fill opacity=1 ]  (269.6,251.7) .. controls (269.6,250.04) and (270.94,248.7) .. (272.6,248.7) -- (338.6,248.7) .. controls (340.26,248.7) and (341.6,250.04) .. (341.6,251.7) -- (341.6,320.7) .. controls (341.6,322.36) and (340.26,323.7) .. (338.6,323.7) -- (272.6,323.7) .. controls (270.94,323.7) and (269.6,322.36) .. (269.6,320.7) -- cycle  ;
\draw (305.6,286.2) node  [font=\huge]  {$\accentset{( -2)}{[ P_{+}]}$};
\draw  [fill={rgb, 255:red, 217; green, 219; blue, 241 }  ,fill opacity=1 ]  (264.1,417.7) .. controls (264.1,416.04) and (265.44,414.7) .. (267.1,414.7) -- (344.1,414.7) .. controls (345.76,414.7) and (347.1,416.04) .. (347.1,417.7) -- (347.1,494.7) .. controls (347.1,496.36) and (345.76,497.7) .. (344.1,497.7) -- (267.1,497.7) .. controls (265.44,497.7) and (264.1,496.36) .. (264.1,494.7) -- cycle  ;
\draw (305.6,456.2) node  [font=\huge]  {$\accentset{( -1)}{[ G]}_{0a}$};
\draw  [fill={rgb, 255:red, 182; green, 226; blue, 211 }  ,fill opacity=1 ]  (395.1,591.7) .. controls (395.1,590.04) and (396.44,588.7) .. (398.1,588.7) -- (463.1,588.7) .. controls (464.76,588.7) and (466.1,590.04) .. (466.1,591.7) -- (466.1,660.7) .. controls (466.1,662.36) and (464.76,663.7) .. (463.1,663.7) -- (398.1,663.7) .. controls (396.44,663.7) and (395.1,662.36) .. (395.1,660.7) -- cycle  ;
\draw (430.6,626.2) node  [font=\huge]  {$\accentset{( 0)}{[ P_{-}]}$};
\draw  [fill={rgb, 255:red, 184; green, 233; blue, 134 }  ,fill opacity=0.74 ]  (130.1,587.7) .. controls (130.1,586.04) and (131.44,584.7) .. (133.1,584.7) -- (228.1,584.7) .. controls (229.76,584.7) and (231.1,586.04) .. (231.1,587.7) -- (231.1,664.7) .. controls (231.1,666.36) and (229.76,667.7) .. (228.1,667.7) -- (133.1,667.7) .. controls (131.44,667.7) and (130.1,666.36) .. (130.1,664.7) -- cycle  ;
\draw (180.6,626.2) node  [font=\huge]  {$\accentset{( 0)}{[ G]}_{\{ab\}}$};
\draw [line width=1.5]    (305.6,323.7) -- (305.6,410.7) ;
\draw [shift={(305.6,414.7)}, rotate = 270] [fill={rgb, 255:red, 0; green, 0; blue, 0 }  ][line width=0.08]  [draw opacity=0] (20.36,-9.78) -- (0,0) -- (20.36,9.78) -- (13.52,0) -- cycle    ;
\draw [line width=1.5]    (275.09,497.7) -- (213.48,581.48) ;
\draw [shift={(211.11,584.7)}, rotate = 306.33] [fill={rgb, 255:red, 0; green, 0; blue, 0 }  ][line width=0.08]  [draw opacity=0] (20.36,-9.78) -- (0,0) -- (20.36,9.78) -- (13.52,0) -- cycle    ;
\draw [line width=1.5]    (336.11,497.7) -- (400.66,585.48) ;
\draw [shift={(403.03,588.7)}, rotate = 233.67] [fill={rgb, 255:red, 0; green, 0; blue, 0 }  ][line width=0.08]  [draw opacity=0] (20.36,-9.78) -- (0,0) -- (20.36,9.78) -- (13.52,0) -- cycle    ;

\end{tikzpicture}
}
\vspace{0.5cm}
\caption{We show how the multiplet of non-relativistic equations of motion describing the quadratic Ricci scalar Newton-Cartan gravity transforms under Galilean boosts.}
\label{fig:diagram3}
\end{figure}
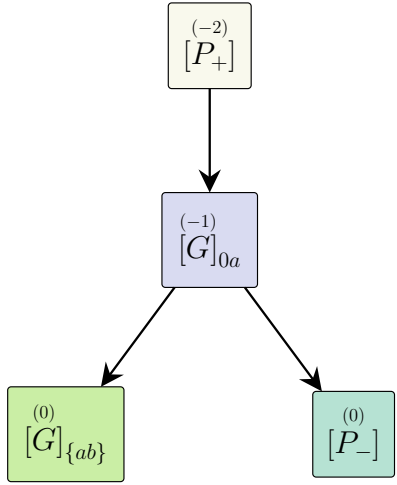

\noindent We have shown that for the Gauss-Bonnet theory it is possible to get the Poisson equation by taking the limit of the equations of motion by deforming the model with the help of a scalar field. We now aim to further extend the analysis. We consider the quadratic Ricci scalar theory, $\mathcal{L}_{\CA}$, whose action we report here for convenience
\begin{align}
\mathcal{S}=&\,  \int\,d^{D}x\, \sqrt{-g} \bigg[\R-\frac{1}{4}F_{\mu\nu}F^{\mu\nu}+\CA \bigg(\R-\frac{1}{4}F_{\mu\nu}F^{\mu\nu}\bigg)^{2}\bigg]\,.
\end{align}
The relativistic equations of motion in this case are:
\begin{subequations}
\begin{align}
[A]^{\mu}:=&\,  \nabla_{\nu}\bigg[F^{\nu\mu}+2\alpha F^{\nu\mu}\bigg(\R-\frac{1}{4}F^{2}\bigg)\bigg]=0\,,\\
\nonumber\\
[G]_{\mu\nu}:=&\,  \Ric_{\mu\nu}-\frac{1}{2}g_{\mu\nu}\R-\frac{1}{2}F_{\mu\rho}F_{\nu}{}^{\rho}+\frac{1}{8}g_{\mu\nu}F^{2}+\nonumber\\
&+2\alpha \bigg(\R-\frac{1}{4}F^{2}\bigg)\bigg(\Ric_{\mu\nu}-\frac{1}{2}F_{\mu\rho}F_{\nu}{}^{\rho}\bigg)-2\alpha\nabla_{(\mu}\nabla_{\nu)}\bigg(\R-\frac{1}{4}F^{2}\bigg)+\nonumber\\
&+\alpha g_{\mu\nu}\bigg[-\frac{1}{2}\bigg(\R-\frac{1}{4}F^{2}\bigg)^{2}+2\nabla_{\rho}\nabla^{\rho}\bigg(\R-\frac{1}{4}F^{2}\bigg)\bigg]=0\,.
\end{align}
\end{subequations}
We consider the same combination \autoref{eq:PoissonCombination} of the two-derivative cases. Expanding we get:
\begin{subequations}
\begin{align}
\accentset{(2)}{[P]}=&\,  \frac{1}{4}(\PC-2\PB)t_{ab}t^{ab}\Big[1+4\Ct_{c0}{}^{c}+2\RicNR_{c}{}^{c}-3t_{0c}t_{0}{}^{c}-2f^{cd}t_{cd}\Big]\,,\\
\nonumber\\
\accentset{(0)}{[P]}=&\,  \frac{1}{2}(3\PA-D\PA+\PC)\RicNR_{a}{}^{a}+\frac{1}{2}(5\PA-D\PA+\PC)\alpha(\RicNR_{a}{}^{a})^2+\nonumber\\
&+2(D\PA-2\PA-\PC)h^{\mu\nu}\nablaNR_{\mu}\nablaNR_{\nu}\RicNR_{a}{}^{a}+...\,,\label{eq:PoissonRSquared}
\end{align}
\end{subequations}
where the dots denote terms proportional to $t_{\mu\nu}$. It is clear  that the order $\cc{2}$ is harmless because of the constraint 
$F_{\mu\nu}=0$, that shifts it to the order $\cc{-2}$.
However, from \autoref{eq:PoissonRSquared} we realize that, at order $\cc{0}$, we do not only get terms proportional to the intrinsic torsion, which can be eliminated using the constraint, but also terms insensitive to it. Just as in Gauss-Bonnet gravity, in this case it is not possible either to cancel all the contributions by choosing the constants $\PA$ and $\PC$ appropriately. To cancel the zero-order term in the expansion we can supplement the system with a further constraint,  
\begin{align}
\langle C \rangle :=\R^{2}+2\nabla_{\mu}\nabla^{\mu}\R=&\,  0\,,
\end{align}
and choose at the same time
\begin{align}
\PC=D\PA-3\PA\,.
\end{align}
In this way, we can make \autoref{eq:PoissonRSquared} vanish. The two constraints after the limit become:
\begin{subequations}
\begin{align}
\RH_{\mu\nu}=&\,  0\,,\\
\accentset{(0)}{\langle C\rangle}:=\RicJ^{2}-2h^{\mu\nu}\nablaNR_{\mu}\nablaNR_{\nu}\RicJ=&\,  0\,,
\end{align}
where the number above recalls at which order of the expansion of the corresponding relativistic constraints the final result appears. Then the Poisson equation is:
 \begin{align}
\accentset{(-2)}{[P_{+}]}:=&\,  (D-2)\PA\RG_{0a}{}^{a}(1-2\CA\RicJ)+2(D-2)\CA\PA\tau^{\mu}\tau^{\nu}\nablaNR_{\mu}\partial_{\nu}\RicJ=0\,. 
\end{align}
In the next section we derive the full set of equations of motion and the corresponding transformations under boost. We will drop the factor $\CA$ and the overall factor $(D-2)\PA$ from the Poisson equations.

\subsubsection{Full Set of Equations of Motion}\label{sec:NRRREOM}
\noindent The full set of non-relativistic equations is:
\begin{align}
\accentset{(-2)}{[P_{+}]}:=&\,  \RG_{0a}{}^{a}(1-2\RicJ)+2\tau^{\mu}\tau^{\nu}\nablaNR_{\mu}\partial_{\nu}\RicJ=0\,,\\
\nonumber\\
\accentset{(0)}{[P_{-}]}:=&\,  \RicJ(1-3\RicJ)=0\,,\\
\nonumber\\
\accentset{(-1)}{[G]}_{0a}:=&\,  -\RicJ_{0a}+2\RicJ_{0a}\RicJ+2\tau^{(\mu}e^{\nu)}{}_{a}\nablaNR_{\mu}\partial_{\nu}\RicJ=0\,,\\
\nonumber\\
\accentset{(0)}{[G]}_{\{ab\}}:=&\,  -\RicJ_{\{ab\}}(1-2\RicJ)+2e^{\mu}{}_{\{a}e^{\nu}{}_{b\}}\nablaNR_{\mu}\partial_{\nu}\RicJ=0\,,
\end{align}
\end{subequations}
where $\accentset{(0)}{[P_{-}]}$ has been obtained by taking the limit of the combination 
\begin{align*}
[P_{-}]:=E^{\mu}{}_{a}E^{\nu a}[G]_{\mu\nu}+(D-3) E^{\mu}{}_{0}E^{\nu}{}_{0}\, [G]_{\mu\nu}\,,
\end{align*}
and the other equations are defined as in \autoref{sec:NRGBEOM}. These should be supplemented with the constraints:
\begin{subequations}
\begin{align}
\RH_{\mu\nu}=&\,  0\,,\\
\accentset{(0)}{\langle C\rangle}:=\RicJ^{2}-2h^{\mu\nu}\nablaNR_{\mu}\partial_{\nu}\RicJ=&\,  0\,.
\end{align}
\end{subequations}
The set of non-relativistic equations and constraints transforms under boosts as:
\begin{subequations}
\begin{align}
\delta_{G}\RH_{\mu\nu}=&\,  0\,,\\
\delta_{G}\accentset{(0)}{\langle C\rangle}=&\,  0\,,\\
\delta_{G}\accentset{(-2)}{[P_{+}]}=&\,  -2\boostp^{a}\accentset{(-1)}{[G]}_{0a}\,,\\
\delta_{G}\accentset{(0)}{[P_{-}]}=&\,  0\,,\\
\delta_{G}\accentset{(-1)}{[G]}_{0a}=&\,  -\boostp^{b}\accentset{(0)}{[G]}_{\{ab\}}+\frac{\boostp_{a}}{D-1}\accentset{(0)}{[P_{-}]}\,,\\
\delta_{G}\accentset{(0)}{[G]}_{\{ab\}}=&\,  0\,.
\end{align}
\end{subequations}
These transformations have to be intended up to the constraints themselves. The action of the boost transformations is visualized in \autoref{fig:diagram3}. The set of equations define a reducible indecomposable representation of the Bargmann algebra. All the equations, except the Poisson equation, can be obtained from the corresponding non-relativistic action.

\FloatBarrier

\section*{Conclusion}\addcontentsline{toc}{section}{\protect\numberline{}Conclusion}\label{sec:Conclusion}
\noindent The present research marks a step forward in understanding higher-derivative corrections to Newton-Cartan gravity. In \autoref{sec:RelaHDT} we introduced the four-derivative theories whose non-relativistic limit has been investigated in this work. In \autoref{sec:NRActionLimit} we studied how to take the non-relativistic limit of quadratic gravity theories, at the level of the action, by introducing a 1-form gauge field and coupling it to the gravity sector. This is a generalization of the procedure used to obtain Newton-Cartan gravity from General Relativity via a limit \cite{Bergshoeff:2015uaa}. We demonstrated that, to avoid divergences in the action and get the magnetic limit, it is sufficient to use a single vector gauge field. The non-relativistic actions have been derived and rewritten in terms of geometric quantities. We provided a general recipe for extending the procedure to higher-order gravity theories.  We also show that higher-order Maxwell terms and their couplings to the gravity sector, used to cancel the divergences in the action, can be obtained from a purely gravitational higher-dimensional theory via Kaluza-Klein reduction. \Autoref{sec:EomLimit} is devoted to the study of the non-relativistic limit of the equations of motion, with special focus on the Poisson equation.  We first review the two-derivative case and identify the main obstacles in obtaining the Poisson equations with higher-derivative corrections. We then considered two specific cases, the Einstein-Gauss-Bonnet theory in \autoref{sec:GBcase} and the quadratic Ricci scalar theory in \autoref{sec:HigherOrderEOM} . In the first case, we derive the higher-order corrected Poisson equation by modifying the initial model with a scalar field. In the second case, to derive the non-relativistic Poisson equation, we supplement the relativistic system with a further on-shell constraint. We also derived, in both cases, the full set of non-relativistic equations of motion. The results define zero-torsion Gauss-Bonnet Newton-Cartan and quadratic Ricci scalar Newton-Cartan gravity theories.

Higher-derivative theories can be affected by Ostrogradsky's Instability Theorem. This can undermine their physical robustness, especially in the context of quantum field theory, due to the appearance of negative energy states. Though they are also often renormalisable, making them appealing in quantum gravity. Therefore, it would be particularly interesting to study in detail the Gauss-Bonnet and quadratic Ricci Newton-Cartan gravity theories and search for their spectrum of solutions.

The present study lays a foundation for a wide array of future investigations, among which we mention:
\begin{itemize}
\item extending the limit to different foliations, by coupling higher-order theories to higher-rank differential forms. This is particularly interesting in the context of string theory and supergravity \cite{Bergshoeff:2025uut, Bergshoeff:2025grj, Bergshoeff:2024nin, Bergshoeff:2023fcf, Bergshoeff:2023ogz,Bergshoeff:2023rkk, Bergshoeff:2021bmc, Bergshoeff:2021tfn}.
\item Investigating whether coupling the scalar field to the gravity sector can be used to obtain the Poisson equation without using constraints. Closely related is the emergence of a local dilatation symmetry.
\item Classifying alternative constraints that yield the Poisson equations from the limit of the equations of motion. These could result in a generalisation of the zero-torsion constraint with higher-order corrections.
\item Deepen the analysis of the higher-dimensional origin of the theories admitting a non-relativistic limit at the level of the action and investigate whether this is based on hidden symmetries \cite{Ciafardini:2024ujx}.
\end{itemize}

Further investigations and generalisations are currently underway and will be reported in due course.

\section*{Acknowledgements}

\noindent We would like to thank José Juan Fernández Melgarejo for his contribution to the first part of the project and the useful discussions. We would also like to thank Eric Bergshoeff, Jan Rosseel and Tomas Ortín for useful discussions. B.C. has been supported by a Margarita Salas Postdoctoral Research Fellowship funded by Next Generation EU. The work of L.R. has been supported in part by the Ramon y Cajal fellowship RYC2023-042671-I , funded by MCIU/AEI/10.13039/501100011033 and FSE+ and in part by the MCI, AEI, FEDER (UE) grant PID2021-125700NAC22.

\appendix
\section{Notation and Conventions}\label{sec:Notation}
\noindent In this work we have adopted the following notation:
\FloatBarrier
\begin{table}[!ht]
\centering
\renewcommand{\arraystretch}{1.5}
\begin{center}
\begin{tabular}{lcl}
\bf {Index}&&\bf{Definition \& values}\\
$M,N,P,Q,...$&&$D+1$-dim curved index ($M=\{\mu,x\}$ or $M=\{\mu,z\}$)\\
$\mu,\nu,\rho,...$&& $D$-dim curved index\\
$\hat{A},\hat{B},\hat{C},...$&&$D$-dim flat index ($\hat{A}=\{0,a\}$)\\
$a ,b ,c,...$&&Transverse flat index $a=1,...,D-1$\\
$x$&&Spatial Isometry direction \\
$z$&&Null Isometry direction
\end{tabular}
\end{center}
\end{table}
\FloatBarrier
\noindent We also use the following definitions:
\begin{subequations}
\begin{align}
t_{\mu\nu}{}=&\,  2\partial_{[\mu}\tau_{\nu]}\,,&z_{\mu\nu}{}^{a}=&\,  2\partial_{[\mu}e_{\nu]}{}^{a}\,,\\
h^{\mu\nu}=&\,  e^{\mu a}e^{\nu}{}_{a}\,,&h_{\mu\nu}=&\,  e_{\mu a}e_{\nu}{}^{a}\,,\\
\tau^{\mu\nu}=&\,  \tau^{\mu}\tau^{\nu}\,,&\tau_{\mu\nu}=&\,  \tau_{\mu}\tau_{\nu}\,,
\end{align}
\end{subequations}
and 
\begin{subequations}
\begin{align}
\Ct_{\mu\nu\rho}=&\,  \nablaNR_{\mu}t_{\nu\rho}\,,\\
\Cm_{\mu\nu\rho}=&\,  \nablaNR_{\mu}f_{\nu\rho}\,,\\
\Cz_{\mu\nu\rho}{}^{a}=&\,  \nablaNR_{\mu}z_{\nu\rho}{}^{a}\,,
\end{align}
\end{subequations}
where the covariant derivatives only contain the non-relativistic connection \autoref{eq:NRConnection}. Furthermore, given a generic non-relativistic tensor $b_{\mu\nu\rho}$ (i.e.~after taking the limit $\cc{}\rightarrow \infty$) we  project curved indices to flat indices as follows
\begin{align}
b_{a0\rho}=&\,  e^{\mu}{}_{a}\tau^{\nu}b_{\mu\nu\rho}\,.
\end{align}
We have also introduced the following variables/couplings:
\FloatBarrier
\begin{table}[!ht]
\centering
\renewcommand{\arraystretch}{1.5}
\begin{center}
\begin{tabular}{lclcc}
\bf {Variables}&&\bf{Definition}&&\bf{Defining Equation}\\
$\CA,\CB,\CC$&&Action couplings&&\autoref{eq:quadraticgravity}\\
$\BA,\BB$&&Dilatation couplings&&\autoref{eq:DilatonTrick1}\\
$\PA,\PB,\PC,\PD$&&Poisson combination coefficients &&\autoref{eq:PoissonCombination},\autoref{eq:PoissonCombinationRevised}\\
\end{tabular}
\end{center}
\end{table}

\FloatBarrier

\section{Definitions \& Useful Relations}\label{sec:UsRel}
\noindent The Riemann tensor is defined by
\begin{align}
\Riem^{\rho}{}_{\sigma\mu\nu}=&\,  \partial_{\mu}\Gamma_{\nu\sigma}^{\rho}-\partial_{\nu}\Gamma_{\mu\sigma}^{\rho}+\Gamma_{\mu\lambda}^{\rho}\Gamma_{\nu\sigma}^{\lambda}-\Gamma_{\nu\lambda}^{\rho}\Gamma_{\mu\sigma}^{\lambda}
\end{align}
with $\Gamma^{\rho}_{\mu\nu}$ the Christoffel symbols,
\begin{align}
\Gamma_{\mu\nu}^{\rho}=&\,  \frac{1}{2}g^{\rho\sigma}(\partial_{\mu}g_{\nu\sigma}+\partial_{\nu}g_{\mu\sigma}-\partial_{\sigma}g_{\mu\nu})\,.
\end{align}
The Ricci tensor and scalar are defined by
\begin{subequations}
\begin{align}
\Ric_{\mu\nu}=&\,  \Riem^{\rho}{}_{\mu\rho\nu}\,,\\
\R=&\,  g^{\mu\nu}\Ric_{\mu\nu}\,.
\end{align}
\end{subequations}
We report in this section relations that have been used in deriving the relativistic equations of motion and manipulating them
\begin{subequations}
\begin{align}
\delta\sqrt{-g}=&\,  -\frac{1}{2}\sqrt{-g}g_{\mu\nu}\delta g^{\mu\nu}\,,\\
\delta \Riem^{\rho}{}_{\sigma\mu\nu}=&\,  \nabla_{\mu}(\delta\Gamma^{\rho}_{\nu\sigma})-\nabla_{\nu}(\delta\Gamma^{\rho}_{\mu\sigma})\,,\\
\delta \Ric_{\mu\nu}=&\,  \nabla_{\rho}(\delta\Gamma^{\rho}_{\nu\mu})-\nabla_{\nu}(\delta\Gamma^{\rho}_{\rho\mu})\,,\\
\delta \R=&\,  \Ric_{\mu\nu}\delta g^{\mu\nu}+\nabla_{\sigma}(g^{\mu\nu}\delta\Gamma^{\sigma}_{\nu\mu}-g^{\mu\sigma}\delta\Gamma^{\rho}_{\rho\mu})\,,\\
[\nabla_{\alpha},\nabla_{\beta}]F_{\mu\nu}=&\,  -\Riem^{\rho}{}_{\mu\alpha\beta}F_{\rho\nu}-\R^{\rho}{}_{\nu\alpha\beta}F_{\mu\rho}\,.
\end{align}
\end{subequations}
The following relations are also useful:  
\begin{subequations}
\begin{align}
\nabla^{\mu} \Ric_{\mu\nu}=&\,  \frac{1}{2}\nabla_{\nu}\R\,,\\
\nabla_{\alpha}F^{\mu\nu}\nabla_{\mu}F_{\beta\nu}=&\,  -\frac{1}{2}\nabla_{\alpha}F_{\mu\nu}\nabla_{\beta}F^{\mu\nu}\,,\\
\nabla_{\mu}F_{\nu\rho}\nabla^{\rho}F^{\mu\nu}=&\,  \frac{1}{2}\nabla_{\mu}F_{\nu\rho}\nabla^{\mu}F^{\nu\rho}\,,\\
\Riem_{\alpha}{}^{\mu\nu\rho}\Riem_{\beta\nu\mu\rho}=&\,  \frac{1}{2}\Riem_{\alpha}{}^{\mu\nu\rho}\Riem_{\beta\mu\nu\rho}\,,\\
F^{\mu\nu}\Riem_{\alpha\mu\beta\nu}=&\,  \frac{1}{2}F^{\mu\nu}\Riem_{\alpha\beta\mu\nu}\,.
\end{align}
\end{subequations}
In analogy with the relativistic case we define the following non-relativistic tensors:
\begin{align}
\RiemNR^{\rho}{}_{\sigma \mu \nu}=&\,  \partial_{\mu}\GammaNR^{\rho}_{\nu\sigma}- \partial_{\nu}\GammaNR^{\rho}_{\mu\sigma}+\GammaNR^{\rho}_{\mu\xi}\GammaNR^{\xi}_{\nu\sigma}-\GammaNR^{\rho}_{\nu\xi}\GammaNR^{\xi}_{\mu\sigma}\,,\\
\RicNR_{\mu \nu}=&\,  \RiemNR^{\rho}{}_{\mu \rho \nu}\,.
\end{align}
Beyond the formal analogy, useful to organize the calculations, these non-relativistic tensors do not transform covariantly under Galilean boosts. The proper definition of geometric non-relativistic curvatures is discussed in \autoref{sec:Geometry}.

\subsection{Non-Relativistic Curvatures Bianchi  Identities}\label{sec:NRCurvaturesIds}
\noindent A set of non-relativistic Bianchi identities is:
\begin{subequations}
\begin{align}
\nablaNR_{[\mu}t_{\nu\rho]}=&\,  -\tau^{\alpha}t_{\alpha[\mu}t_{\nu\rho]}\,,\\
\nablaNR_{[\mu}z_{\nu\rho]}{}^{a}=&\,  -\tau^{\alpha}z_{\alpha[\mu}{}^{a}t_{\nu\rho]}\,,\\
\nablaNR_{[\mu}f_{\nu\rho]}=&\,  -\tau^{\alpha}f_{\alpha[\mu}t_{\nu\rho]}\,,\\
\RiemNR^{\alpha}{}_{[\mu\nu\rho]}=&\,  -\frac{1}{2}\tau^{\sigma}e^{\alpha}{}_{a}t_{[\mu\nu}z_{\rho]\sigma}{}^{a}+\frac{1}{2}\tau^{\sigma} h^{\alpha\xi}z_{\sigma\xi a}e_{[\mu}{}^{a}t_{\nu\rho]}+\frac{1}{2}h^{\alpha\xi}t_{[\mu\nu}f_{\rho]\xi}+\nonumber\\
&+\frac{1}{2}h^{\alpha\xi}f_{0\xi}\tau_{[\mu}t_{\nu\rho]}\,.
\end{align}
\end{subequations}
Further useful identities are the following:
\begin{subequations}
\begin{align}
\tau_{\sigma}\RiemNR^{\sigma}{}_{\mu\nu\rho}=&\,  0\,,\\
\RicNR_{[ab]}=&\,  -f_{[a}{}^{c}t_{b]c}-t_{ab}z_{0c}{}^{c}+t_{[a}{}^{c}z_{0b]c}+z_{0c[b}t_{a]}{}^{c}\,,\\
\RiemNR_{abcd}=&\,  -\RiemNR_{bacd}\,,\\
\RiemNR^{\sigma}{}_{\rho\mu\nu}=&\,  -\RiemNR^{\sigma}{}_{\rho\nu\mu}\,,\\
\RiemNR_{abcd}=&\,  \RiemNR_{cdab}+\frac{1}{2}(t_{ab}f_{cd}-t_{cd}f_{ab})
-t_{ac}z_{0(bd)}+t_{ad}z_{0(bc)}+t_{bc}z_{0(ad)}-t_{bd}z_{0(ac)}\,,\\
3\RicNR_{a[bcd]}=&\,  \frac{3}{2}(z_{0a[b}t_{cd]}+t_{[bc|}z_{0|d]a}-f_{a[b}t_{cd]})\,.
\end{align}
\end{subequations}

\section{On Newton-Cartan Geometry}\label{sec:Geometry}
\noindent In this section we introduce the basics of the Newton-Cartan geometry to rewrite our results in terms of covariant quantities. We have defined the non-relativistic connection $\GammaNR_{\mu\nu}^{\rho}$ in \autoref{eq:NRConnection} such that the non-relativistic analogue of metric compatibility relations, $\nablaNR_{\mu}\tau_{\nu}= 0 = \nablaNR_{\mu}h^{\nu\rho}$, are satisfied. We can now introduce a boost and transverse rotations spin connections, $\oG_{\mu}{}^{a}$ and $\oJ_{\mu}{}^{a b}$ respectively, following \cite{Bergshoeff:2022fzb}, by considering the analogue of the relativistic vielbein postulate:
\begin{align}
\partial_{\mu}e_{\nu}{}^{a}-\oJ_{\mu}{}^{a}{}_{b}e_{\nu}{}^{b}-\oG_{\mu}{}^{a}\tau_{\nu}-\GammaNR_{\mu\nu}^{\rho}e_{\rho}{}^{a}=&\,  0\,.\label{eq:NRVielbeinPostulate}
\end{align}
Solving for the two spin connections we get
\begin{subequations}
\begin{align}
\oG_{\mu}{}^{a}=&\,  \tau^{\nu}(\partial_{\mu}e_{\nu}{}^{a}-\GammaNR_{\mu\nu}^{\rho}e_{\rho}{}^{a})\,,\\
\oJ_{\mu}{}^{a b}=&\,  e^{\nu b}(\partial_{\mu}e_{\nu}{}^{a}-\GammaNR_{\mu\nu}^{\rho}e_{\rho}{}^{a})\,.
\end{align}
\end{subequations}
Note that the transverse rotations connection is antisymmetric in the flat indices, although in the expression above it is not manifest. The connections just defined transform as follows under $U(1)$ gauge symmetry, Galilean boost and transverse rotations:
\begin{subequations}
\begin{align}
\delta \GammaNR_{\mu\nu}^{\rho}=&\,  \frac{1}{2}\boostp_{a}h^{\rho\sigma}(e_{\nu}{}^{a}t_{\mu\sigma}+e_{\mu}{}^{a}t_{\nu\sigma}-e_{\sigma}{}^{a}t_{\mu\nu})\,,\\
\delta \oG_{\mu}{}^{a}=&\,  \partial_{\mu}\boostp^{a}-\boostp_{b}\oJ_{\mu}{}^{a b}+\lorentzp^{a}{}_{b}\oG_{\mu}{}^{b}-\frac{1}{2}
\boostp_{b}e_{\mu}{}^{b}t_{0}{}^{a}+\frac{1}{2}\boostp^{a}t_{\mu 0}\,,\\
\delta\oJ_{\mu}{}^{a b}=&\,  \partial_{\mu}\lorentzp^{ab}+2\lorentzp^{[b}{}_{c}\oJ_{\mu}{}^{a]c}+\frac{1}{2}\boostp_{c}e_{\mu}{}^{c}t^{a b}+t_{\mu[b}\boostp_{a]}\,.
\end{align}
\end{subequations}
We can define the following covariant curvatures:
\begin{subequations}
\begin{align}
\RH_{\mu\nu}=&\,  t_{\mu\nu}\,,\\
\RP_{\mu\nu}{}^{a}=&\,  z_{\mu\nu}{}^{a}-2\oJ_{[\mu}{}^{ab}e_{\nu]b}-2\oG_{[\mu}{}^{a}\tau_{\nu]}\,,\\
\RM_{\mu\nu}=&\,  f_{\mu\nu}+2\oG_{[\mu}{}^{a}e_{\nu]a}\,,\\
\RG_{\mu\nu}{}^{a}=&\,  2\partial_{[\mu}\oG_{\nu]}{}^{a}-2\oJ_{[\mu}{}^{a b}\oG_{\nu] b}-\oG_{[\nu}{}^{b}e_{\mu] b}t_{0}{}^{a}+\oG_{[\nu}{}^{a}t_{\mu]0}\,,\\
\RJ_{\mu\nu}{}^{ab}=&\,  2\partial_{[\mu}\oJ_{\nu]}{}^{a b}+2\oJ_{[\mu}{}^{a c}\oJ_{\nu]}{}^{b}{}_{c} -2\oG_{[\mu}{}^{[a}t_{\nu]}{}^{b]}-\oG_{[\mu}{}^{c}e_{\nu] c}t^{ab}\,.
\end{align}
\end{subequations}
$\RM_{\mu\nu}$ and $\RP_{\mu\nu}{}^{a}$ vanish once the expression of the spin connection is plugged into their definitions. 
The curvatures transform as
\begin{subequations}
\begin{align}
\delta \RH_{\mu\nu}=&\,  0\,,\\
\delta \RP_{\mu\nu}{}^{a}=&\,  \lorentzp^{a}{}_{b}\RP_{\mu\nu}{}^{b}\,,\\
\delta \RM_{\mu\nu}=&\,  -\boostp_{a}\RP_{\mu\nu}{}^{a}\,,\\
\delta \RG_{\mu\nu}{}^{a}=&\,  \lorentzp^{a}{}_{b}\RG_{\mu\nu}{}^{b}
-\boostp_{b}\RJ_{\mu\nu}{}^{a b}-\frac{1}{2}\boostp^{a}\tau^{\rho}\nablaNR_{\rho}\RH_{\mu\nu}+\nonumber\\
&+\boostp_{b}\tau^{\rho}e^{\sigma a}e_{[\mu}{}^{b}\nablaNR_{\nu]}\RH_{\rho\sigma}+\frac{1}{2}\boostp_{b}e_{[\mu}{}^{b}\RH_{\nu]0}\RH_{0}{}^{a}\,,\\
\delta \RJ_{\mu\nu}{}^{a b}=&\,  -2\lorentzp^{[a}{}_{c}\RJ_{\mu\nu}{}^{b]c}++e^{\rho[a}\boostp^{b]}\nablaNR_{\rho}\RH_{\mu\nu}+\nonumber\\
&+\frac{2}{3}\boostp^{[a}\RH_{0}{}^{b]}\RH_{\mu\nu}+\frac{1}{3}\boostp^{[a}\RH_{[\mu}{}^{b]}\RH_{\nu]0}+\nonumber\\
&+\boostp_{c}e^{\rho a}e^{\sigma b}e_{[\nu}{}^{c}\nablaNR_{\mu]}\RH_{\rho\sigma}-\boostp_{c}e_{[\mu}{}^{c}\RH_{\nu]}{}^{[a}\RH_{0}{}^{b]}\,.
\end{align}
\end{subequations}
We define 
\begin{align}
\RicJ_{a b}=&\,  \RJ_{a c b}{}^{c}\,,
\end{align}
which is symmetric in its indices and transforms as
\begin{align}
\delta \RicJ_{a b}=&\,  2\lorentzp^{(a}{}_{c}\RicJ^{b)}{}_{c}-\boostp^{(a}\CDRH_{d}{}^{b)d}+\boostp_{d}\CDRH^{(ab)d}+\nonumber\\
&+\frac{1}{2}\boostp^{(a}\RH^{b)c}\RH_{0c}-\frac{1}{2}\boostp_{c}\RH_{0}{}^{(a}\RH^{b)c}\,.
\end{align}
The definition above induces the definition
\begin{align}
\RicJ=&\,  \RicJ_{a}{}^{a}=\RJ_{a b}{}^{a b}\,,
\end{align}
transforming as
\begin{align}
\delta \RicJ=&\,  -2\boostp_{a}\CDRH^{ba}{}_{b}+\boostp_{a}\RH_{0b}\RH^{ab}\,.
\end{align}
We also have the following Bianchi identity:
\begin{subequations}
\begin{align}
\nablaNR_{[\mu}\RH_{\nu\rho]}-\tau^{\xi}\RH_{[\mu\nu}\RH_{\rho]\xi}=&\,  0\,,
\end{align}
\end{subequations}
and 
\begin{subequations}
\begin{align}
\RJ_{abcd}=&\,  \RJ_{cdab}\,,\\
\RJ_{a[bcd]}=&\,  0\,,\\
\RG_{ba}{}^{b}=&\,  \RJ_{0ba}{}^{b}=\RicJ_{0a}\,,\\
2\RJ_{0[ab]c}=&\,  \RG_{abc}\,,\\
\RJ_{0[abc]}=&\,  \RG_{[abc]}=0\,.
\end{align}
\end{subequations}
The expression of the non-relativistic tensor $\RicNR_{\mu\nu}$ in terms of the geometric quantities introduced in this section is the following:
\begin{align}
\RicNR_{\mu\nu}=&\,  \tau_{\mu}\RG_{\nu a}{}^{a}-e_{\mu}{}^{a}\RJ_{\nu b a}{}^{b}-\oG_{\nu}{}^{a}\RH_{\mu a}+\frac{1}{2}e^{\rho}{}_{a}\oG_{\rho}{}^{a}\RH_{\mu\nu}+\nonumber \\
&+e^{\rho b}\oG_{\rho a}e_{(\mu}{}^{a}\RH_{\nu)b}\,.
\end{align}
It is also useful to report here the transformation rule of $\CDRH_{\mu\nu\rho}$ under all symmetries except diffeomorphisms:
\begin{align}
\delta\CDRH_{\mu\nu\rho}=&\,  \boostp_{a}\Big(e_{[\nu}{}^{a}\RH_{\rho]b}\RH_{\mu}{}^{b}-\RH_{\mu[\nu}\RH_{\rho]}{}^{a}\Big)\,.
\end{align}


\bibliography{bibliography}{}

\appendix
\end{document}